\documentclass[aps,prd,nofootinbib,notitlepage,longbibliography,twocolumn, superscriptaddress,10pt]{revtex4-1}
\usepackage{amsmath,amssymb,amsfonts}

\usepackage[utf8]{inputenc}
\usepackage{hyperref}
\usepackage{graphicx}  

\usepackage{tikz}
\usetikzlibrary{decorations.markings}
\usetikzlibrary{shapes.geometric}

\newcommand{\eq}{\begin{equation}}
\newcommand{\eqx}{\end{equation}}
\newcommand{\eqn}{\begin{eqnarray}}
\newcommand{\eqnx}{\end{eqnarray}}
\newcommand{\f}[2]{\frac{#1}{#2}}

\newcommand{\eps}{\varepsilon}
\newcommand{\dl}{\delta}
\newcommand{\Dl}{\Delta}
\renewcommand{\th}{\theta}
\newcommand{\sg}{\sigma}
\newcommand{\lm}{\lambda}

\DeclareMathOperator{\tr}{tr}
\def\res{\mathop{Res}}

\newcommand{\hb}{\bar{h}}
\newcommand{\zb}{\bar{z}}

\newcommand{\Ab}{\bar{A}}
\newcommand{\ab}{\bar{a}}
\newcommand{\lb}{\bar{l}}
\newcommand{\Tb}{\overline{T}}
\newcommand{\xb}{{\bar{x}}}
\newcommand{\lmb}{{\bar{\lambda}}}
\newcommand{\partialb}{\bar{\partial}}

\newcommand{\cor}[1]{\left\langle{#1}\right\rangle}
\newcommand{\ket}[1]{\left|{#1}\right\rangle}
\newcommand{\bra}[1]{\left\langle{#1}\right|}

\newcommand{\OO}[1]{{\mathcal O}\left(#1\right)}

\newcommand{\NN}{{\mathcal N}}
\newcommand{\GG}{{\mathcal G}}
\newcommand{\RR}{{\mathbb R}}
\newcommand{\CC}{{\mathbb C}}
\newcommand{\ZZ}{{\mathbb Z}}
\newcommand{\PP}{{\mathcal P}}
\newcommand{\DD}{{\mathcal D}}

\newcommand{\qq}{\quad\quad}
\newcommand{\qqqq}{\quad\quad\quad\quad}

\newcommand{\arr}[4]{\left(
\begin{matrix}
{#1} & {#2}\\
{#3} & {#4} \\
\end{matrix}
\right)}

\newcommand{\red}[1]{{#1}}

\tikzset{->-/.style={decoration={
markings,
mark=at position .5 with {\arrow[scale=2]{>}}},postaction={decorate}}}

\tikzset{-<-/.style={decoration={
markings,
mark=at position .5 with {\arrow[scale=2]{<}}},postaction={decorate}}}

\tikzset{-s>-/.style={decoration={
markings,
mark=at position #1 with {\arrow[scale=1.5]{>}}},postaction={decorate}}}

\newcommand{\figrules}{

\raisebox{0.5cm}{a)}\hspace{0.25cm}
\begin{tikzpicture}[scale=1.3]

\node (0) at (-1.5, -0.4) {$\delta(x_f)$};
\node (1) at (1.5, -0.4) {$\frac{1}{x_i^{2h}}$};
\node (2) at (-1.7, 0.2) {$z_f$};
\node (3) at (1.7, 0.2) {$z_i$};
\node (2) at (0, 2.0) {$W(z_f, x_f; z_i, x_i)$};

\draw[-<-, line width=1pt, color=green] (-1.5,0) arc[start angle=180, end angle=0, radius=1.5] (1.5,0);
\draw[line width=2pt] (-2,0) -- (2,0);

\end{tikzpicture}
\hspace{2cm}
\raisebox{0.5cm}{b)}\hspace{0.25cm}
\begin{tikzpicture}[scale=1.3]

\node (0) at (0,-1.2) {};
\node (1) at (0,1.25) {$z$};
\node (2) at (-1.0,-0.2) {$x_1$};
\node (3) at (0,-0.2) {$x_2$};
\node (4) at (1,-0.2) {$x_3$};
\node (5) at (0,-0.8) {$f_{123}(x_1,x_2,x_3)$};

\draw[->-, line width=1pt, color=green] (0,1.0) arc[start angle=105, end angle=160, radius=1.5];
\draw[->-, line width=1pt, color=green] (0,1.0) arc[start angle=90, end angle=90] -- (0,0.0);
\draw[->-, line width=1pt, color=green] (0,1.0) arc[start angle=75, end angle=20, radius=1.5];

\draw[color=green, fill=green] (0,1.0) circle [radius=0.08cm];

\end{tikzpicture}
\hspace{2cm}
\raisebox{0.5cm}{c)}\hspace{0.25cm}
\begin{tikzpicture}[scale=1.3]

\node (0) at (0,-1.2) {};
\node (1) at (0,1.25) {$z$};
\node (2) at (-1.0,-0.2) {$x_1$};
\node (3) at (0,-0.2) {$x_2$};
\node (4) at (1,-0.2) {$x_3$};
\node (5) at (0,-0.8) {$f_{\bar{1}23}(x_1,x_2,x_3)$};

\draw[-<-, line width=1pt, color=green] (0,1.0) arc[start angle=105, end angle=160, radius=1.5];
\draw[->-, line width=1pt, color=green] (0,1.0) arc[start angle=90, end angle=90] -- (0,0.0);
\draw[->-, line width=1pt, color=green] (0,1.0) arc[start angle=75, end angle=20, radius=1.5];

\draw[color=green, fill=green] (0,1.0) circle [radius=0.08cm];

\end{tikzpicture}

}

\newcommand{\qnma}{%
\begin{tikzpicture}[scale=1.3]

\node (A) at (-0.6,-0.8) {};
\node (B) at (0.6,-0.8) {};

\node at (-0.72,-0.95) {$\varepsilon$};
\node at (0.72,-0.95) {$\varepsilon$};

\node[star, fill=red, star point ratio=2, star points=8, inner sep=0.05cm] (0) at (0,0) {};

\draw[-s>-=0.7,line width=1pt, color=green] plot [smooth] coordinates {(A) (0,-0.4) (B)};

\draw[line width=2pt] (0,0) circle [radius=1.0cm];

\end{tikzpicture}
}

\newcommand{\qnmb}{%
\begin{tikzpicture}[scale=1.3]

\node (A) at (-0.6,-0.8) {};
\node (B) at (0.6,-0.8) {};

\node at (-0.72,-0.95) {$\varepsilon$};
\node at (0.72,-0.95) {$\varepsilon$};

\node[star, fill=red, star point ratio=2, star points=8, inner sep=0.05cm] (0) at (0,0) {};

\draw[-s>-=0.7,line width=1pt, color=green] plot [smooth] coordinates {(A) (-0.35,0) (0,0.3) (0.35,0) (B)};

\draw[line width=2pt] (0,0) circle [radius=1.0cm];

\end{tikzpicture}
}

\newcommand{\qnmc}{%
\begin{tikzpicture}[scale=1.3]

\node (A) at (-0.6,-0.8) {};
\node (B) at (0.6,-0.8) {};

\node at (-0.72,-0.95) {$\varepsilon$};
\node at (0.72,-0.95) {$\varepsilon$};

\node[star, fill=red, star point ratio=2, star points=8, inner sep=0.05cm] (0) at (0,0) {};

\draw[-s>-=0.9,line width=1pt, color=green] plot [smooth] coordinates {(A) (-0.35,0) (0,0.3) (0.25,0) (0.1,-0.2) (-0.1,-0.25) (-0.3,0.2) (0,0.45) (0.5,0) (B)};

\draw[line width=2pt] (0,0) circle [radius=1.0cm];

\end{tikzpicture}
}

\newcommand{\plateau}{%
\begin{tikzpicture}[scale=1.3]

\node (A) at (-0.6,-0.8) {};
\node (B) at (0.6,-0.8) {};

\node at (-0.72,-0.95) {$\varepsilon$};
\node at (0.72,-0.95) {$\varepsilon$};

\node[star, fill=red, star point ratio=2, star points=8, inner sep=0.05cm] (0) at (0,0) {};

\draw[->-=0.5,line width=1pt, color=green] plot  coordinates {(A) (-0.4,-0.2)};

\draw[-s>-=0.5,line width=1pt, color=green] plot coordinates {(0.4,-0.2) (B)};

\draw[-s>-=0.8,line width=1pt, color=green] plot [smooth] coordinates {(-0.4,-0.2) (0,-0.4) (0.4,-0.2)};
\draw[-s>-=0.8,line width=1pt, color=green] plot [smooth] coordinates {(0.4,-0.2) (0.3,0.2) (0,0.4) (-0.3,0.2) (-0.4,-0.2)};

\node at (0,-0.6) {$\sigma$};
\node at (0,0.6) {$\sigma$};

\draw[color=green, fill=green] (-0.4,-0.2) circle [radius=0.08cm];
\draw[color=green, fill=green] (0.4,-0.2) circle [radius=0.08cm];

\draw[line width=2pt] (0,0) circle [radius=1.0cm];

\end{tikzpicture}
}

\newcommand{\mirrora}{%
\begin{tikzpicture}[scale=1.3]

\node (A) at (-0.6,-0.8) {};
\node (B) at (0.6,-0.8) {};

\node at (-0.72,-0.95) {$\varepsilon$};
\node at (0.72,-0.95) {$\varepsilon$};

\node[star, fill=red, star point ratio=2, star points=8, inner sep=0.05cm] (0) at (0,0) {};

\draw[-s>-=0.7,line width=1pt, color=green] plot [smooth] coordinates {(A) (0,-0.4) (B)};
\draw[-s>-=0.9,line width=1pt, color=violet] plot [smooth] coordinates {(-0.6,-0.73) (0,-0.34) (0.6,-0.73)};

\draw[line width=2pt] (0,0) circle [radius=1.0cm];

\end{tikzpicture}
}

\newcommand{\mirrorb}{%
\begin{tikzpicture}[scale=1.3]

\node (A) at (-0.6,-0.8) {};
\node (B) at (0.6,-0.8) {};

\node at (-0.72,-0.95) {$\varepsilon$};
\node at (0.72,-0.95) {$\varepsilon$};

\node[star, fill=red, star point ratio=2, star points=8, inner sep=0.05cm] (0) at (0,0) {};

\draw[-s>-=0.9,line width=1pt, color=violet] plot [smooth] coordinates {(A) (-0.35,0) (0,0.3) (0.25,0) (0.1,-0.2) (-0.1,-0.25) (-0.3,0.2) (0,0.45) (0.5,0) (B)};
\draw[-s>-=0.7,line width=1pt, color=green] plot [smooth] coordinates {(A) (0,-0.4) (B)};

\draw[line width=2pt] (0,0) circle [radius=1.0cm];

\end{tikzpicture}
}

\newcommand{\plateaubig}{%
\begin{tikzpicture}[scale=1.3]

\node at (-0.1,-1.15) {};
\node at (0.1,-1.15) {};

\node (A) at (-0.6,-0.8) {};
\node (B) at (0.6,-0.8) {};

\node at (-0.72,-0.95) {$\varepsilon$};
\node at (0.72,-0.95) {$\varepsilon$};

\node[star, fill=red, star point ratio=2, star points=8, inner sep=0.05cm] (0) at (0,0) {};

\draw[-s>-=0.5,line width=1pt, color=violet] plot  coordinates {(A) (-0.4,-0.2)};

\draw[-s>-=0.5,line width=1pt, color=violet] plot  coordinates {(0.4,-0.2) (B)};

\draw[-s>-=0.8,line width=1pt, color=violet] plot [smooth] coordinates {(-0.4,-0.2) (0,-0.4) (0.4,-0.2)};
\draw[-s>-=0.8,line width=1pt, color=violet] plot [smooth] coordinates {(0.4,-0.2) (0.3,0.2) (0,0.4) (-0.3,0.2) (-0.4,-0.2)};

\node at (0,-0.55) {$\sigma$};
\node at (0,0.55) {$\sigma$};

\draw[color=violet, fill=violet] (-0.4,-0.2) circle [radius=0.08cm];
\draw[color=violet, fill=violet] (0.4,-0.2) circle [radius=0.08cm];

\node at (-0.65,-0.2) {$z_1$};
\node at (0.65,-0.2) {$z_2$};

\draw[line width=2pt] (0,0) circle [radius=1.0cm];

\end{tikzpicture}
}

\newcommand{\plateaubigb}{%
\begin{tikzpicture}[scale=1.3]

\node (A) at (-0.1,-1.0) {};
\node (B) at (0.1,-1.0) {};

\node at (-0.1,-1.15) {$\varepsilon$};
\node at (0.1,-1.15) {$\varepsilon$};

\node[star, fill=red, star point ratio=2, star points=8, inner sep=0.05cm] (0) at (0,0) {};

\draw[-s>-=0.5,line width=1pt, color=green] plot  coordinates {(A) (-0.4,-0.2)};

\draw[-s>-=0.5,line width=1pt, color=green] plot  coordinates {(0.4,-0.2) (B)};

\draw[-s>-=0.8,line width=1pt, color=green] plot [smooth] coordinates {(-0.4,-0.2) (0,-0.4) (0.4,-0.2)};
\draw[-s>-=0.8,line width=1pt, color=green] plot [smooth] coordinates {(0.4,-0.2) (0.3,0.2) (0,0.4) (-0.3,0.2) (-0.4,-0.2)};

\node at (0,-0.55) {$\sigma$};
\node at (0,0.55) {$\sigma$};

\draw[color=green, fill=green] (-0.4,-0.2) circle [radius=0.08cm];
\draw[color=green, fill=green] (0.4,-0.2) circle [radius=0.08cm];

\node at (-0.65,-0.2) {$z_1$};
\node at (0.65,-0.2) {$z_2$};

\draw[line width=2pt] (0,0) circle [radius=1.0cm];

\end{tikzpicture}
}

\newcommand{\twop}[5]{%
\raisebox{-0.6cm}{\begin{tikzpicture}

\node[draw, ellipse, minimum height=0.5cm, minimum width=1cm] (A) at (0,0) {#5};
\coordinate (B) at (A.50);
\coordinate (C) at (A.130);

\path (B)++(50:0.57) node (D) {};
\path (C)++(130:0.57) node (E) {};

\path (B)++(50:0.6) node (D2) {#2};
\path (C)++(130:0.6) node (E2) {#1};

\draw (B) -- (D);
\draw (C) -- (E);

\node at (0,0.45) {#3};
\node at (0,-0.45) {#4};

\end{tikzpicture}}
}

\begin{document}

\title{The Ising model as a window on quantum gravity with matter}

\author{Romuald A. Janik}
\email{romuald.janik@gmail.com}
\affiliation{Institute of Theoretical Physics and Mark Kac Center for Complex Systems Research, Jagiellonian University,  ul. {\L}ojasiewicza 11, 30-348 Krak{\'o}w, Poland}

\begin{abstract}
We argue that the Ising model CFT can be used to obtain some clear insights into 3D (quantum) gravity with matter. We review arguments for the existence of its holographic description, and concentrate on the time dependence of perturbations of the theory at high temperature, which would correspond to throwing matter into a black hole in the dual picture. Apart from an expected QNM-like exponential damping, we observe a plateau, a burst and a subsequent re-emergence of the whole signal, the latter being apparently at odds with a black hole interpretation. We provide an explanation of this phenomenon in terms of the properties of bulk matter fields interacting with the BTZ black hole and the fact that the geometry/metric is not fundamental but a derived quantity in the Chern-Simons formulation of 3D gravity. This allows for evading the black hole information paradox in the present context.
\end{abstract}

\maketitle

The AdS/CFT correspondence~\cite{Maldacena:1997re}  posits the equivalence of various quantum field theories with theories incorporating gravity in a higher dimensional spacetime, the quintessential example being the equivalence of $\NN=4$ SYM with string theory on $AdS_5 \times S^5$. It has led to immense insight into the dynamics of strongly coupled gauge theories from various classical gravity or string computations.
One of the most fascinating potential uses of the AdS/CFT correspondence is to explore gravity in its quantum regime using the dual quantum field theory as its nonperturbative quantum definition.
This led to numerous interesting results in the 2D JT gravity/SYK model~\cite{Sachdev:2010um,Mertens:2022irh}, insights into black hole evaporation~\cite{Almheiri:2019hni}, bulk interpretations of conformal blocks in 2D CFTs~\cite{Bhatta:2016hpz,Besken:2016ooo,Fitzpatrick:2016mtp,Besken:2017fsj,Hikida:2017ehf,Hikida:2018dxe,Besken:2018zro} etc. 

One of the difficulties in pursuing this line of investigation is the dichotomy of either having a good control over the boundary QFT or having a semiclassical gravity regime in order to appropriately interpret and understand the field theoretical calculation from a gravitational perspective. 
In the case of 2D CFTs, a semiclassical limit typically involves $c\to \infty$, for which unfortunately we do not have exactly solvable 2D CFTs (unless they possess many additional symmetries which lead to e.g. higher-spin gravity~\cite{Gaberdiel:2010pz}, where the spacetime geometrical interpretation is somewhat less clear).

We should also note in this context, that often the term holographic CFT is used to denote that there is a clear separation between gravitational and nongravitational degrees of freedom~\cite{Heemskerk:2009pn}. Here we use the term holographic description in a much more general manner, \emph{without} requiring any such separation (a prime example being $\NN=4$ SYM at weak coupling). 
On the contrary, the deeply quantum regime of intermingled gravity and matter would be of chief interest for us.

In this paper, we would like to argue that one can obtain some clear insights into a (quantum) 3D gravity+matter system even by studying the simplest exactly solvable 2D CFT --- the Ising model CFT (or similarly other Virasoro minimal model CFTs).

\begin{figure}[t]
    \centering
    \includegraphics[width=\columnwidth]{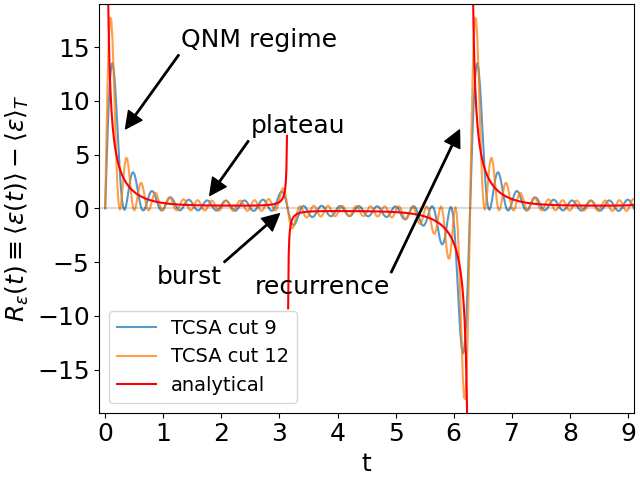}
    \caption{Temporal evolution of a perturbation of the Ising CFT at $T=0.5$ showing the main regimes discussed in the paper --- exponential decay, plateau and a reemergence of the whole signal, as well as a ``burst'' at $t=\pi$. The analytical result from the torus 2-point function is superimposed on TCSA simulations with two different level truncations.}
    \label{fig.tcsa}
\end{figure}

Clearly, as $c=\f{1}{2}$, this theory is apparently in a deeply quantum regime on the dual side. However, we can focus on the high temperature limit (relative to the size of the spatial circle) and then effectively recover some semiclassical behaviour. On the other hand, as we are studying the theory on a cylinder of size $R=2\pi$, all modes have quantized momenta, hence there remains this source of ``quantumness'' also for gravity (dual to the CFT energy-momentum tensor).
As we will see, this will manifest itself in intriguing behaviour for large times, of the order of inverse level spacing.

We adopt the most standard setup of studying a black hole in real time using the AdS/CFT correspondence, by observing some matter falling into the black hole at the linearized level, which would then be described using black hole quasi-normal modes. On the field theory side, this corresponds to perturbing the thermal state (i.e. density matrix) of the theory and observing its temporal evolution.

For the Ising model CFT, this calculation can be done exactly, allowing one to study the exact linearized evolution for arbitrary times. 
We do that explicitly for the $\eps$ operator with $h=\hb=\f{1}{2}$.
We find several distinct phenomena (see Fig.~\ref{fig.tcsa}) starting with the (rather obvious) classical quasi-normal-like exponential decay, which, however, does not decay to zero but reaches a nonzero plateau. The plateau is no longer universal but depends on the details of the theory. Then, after a time of $2\pi$, we observe that the whole signal re-emerges again, completely unchanged. This recurrence is quite at odds with any gravitational black hole interpretation in contrast to the initial QNM-like exponential decay, which fits the black hole picture very nicely.

In this paper, we provide a bulk interpretation of these phenomena in terms of Wilson line networks in the dual gravitational Chern-Simons theory. We are led to conclude that the appearance of the recurrence can be linked to the way that the bulk matter fields as implemented by the Wilson line networks interact with the gravitational system in the Chern-Simons description. 
The key characteristics of this interaction can be read off from the known exact behaviour of the observable in the Ising CFT.
In the Chern-Simons formulation, the spacetime metric and hence the geometry is not fundamental, but rather a derived concept --- opening up the possibility of matter dynamics leading to the recurrence, even in an apparently conventional black hole background, thus essentially sidestepping in this way a variant of the black hole information paradox~\cite{Hawking:1976ra}.
This mechanism would be impossible, however, if the metric was really a fundamental ingredient of the gravity theory relevant in this context.

However, in order to be able to draw any conclusions on gravity and black holes, we have to ascertain
whether a holographical bulk theory is at all applicable to the Ising CFT and more generally to minimal model CFTs. And even if yes, whether we can indeed interpret the dual to a thermal state in Ising CFT at high temperatures as a black hole, given the small value of the central charge $c=\f{1}{2}$.

To this end, we review arguments that the Bershadsky-Ooguri identification~\cite{Bershadsky:1989mf} of hidden $SL(2,\RR)$ symmetry in minimal model CFTs and the corresponding isomorphism of Virasoro representations very strongly indicates that a dual (quantum) Chern-Simons $SL(2,\RR)$ description (with matter fields) exists for these theories. 

Given the above, a holographically inspired heuristic quantization of the Chern-Simons theory realizes both the CS gauge fields as well as the bulk metric as concrete quantum operators in the 2D CFT. 
Then, the thermal expectation value of the bulk metric operator indeed evaluates to the BTZ black hole in the high temperature limit. But surprisingly, at finite temperatures the black hole character of the average geometry is lifted. This effect is, however, not relevant for the recurrence mentioned earlier which occurs even at arbitrarily high temperatures.

Let us finally clarify what do we mean by ``(quantum) gravity'' in the context of the present paper. We identify gravity with the subsector of the dual bulk holographic theory, which corresponds to  the energy-momentum tensor of the boundary quantum field theory. So, in particular, it is not ``pure gravity'' and it is generically accompanied by appropriate bulk matter fields characteristic of the CFT's operator content. 
Along these lines,
we allow for imposing any properties/superselection rules (like restricting the bulk topology) which would be necessary for preserving the holographic duality with the given \emph{single} boundary CFT, even if those requirements would be incompatible with our usual preconceptions on gravity. 
So we would like to ``read off'' what the dual gravity should be like in order for holography to work.

Indeed, there may exist distinct and inequivalent formulations of gravity --- like VirasoroTQFT~\cite{Collier:2023fwi,Collier:2024mgv} versus Chern-Simons, with the former leading e.g. to ensembles of boundary theories~\cite{Chandra_2022} etc. Thus we aim to study here \emph{a} theory of gravity instead of \emph{the} theory of gravity --- without any pretentions for a \emph{unique} theory of gravity.

\red{Before we proceed, let us comment more on the relevance of these considerations to the black hole information paradox~\cite{Hawking:1976ra}. 
We are not dealing here with the classical case of a collapse and evaporation of a black hole, but rather with the temporal evolution of perturbations of a thermal state corresponding to perturbations of an eternal black hole (see~\cite{Maldacena:2001kr} for a similar setup and partial resolution of  the paradox in semiclassical gravity).
The setup considered here exhibits exactly periodic unitary evolution on the CFT side, which is thus in acute tension with the causal structure of the dual black hole spacetime.
As we perform the calculations in an $AdS_3/CFT_2$ context of $c=\f{1}{2}$ Ising CFT, one may worry that the quantum gravity spacetime may be so far away from a semiclassical picture, that the causal intuitions loose their meaning.
We find, however, that at \emph{high temperatures} the calculation of the relevant observables is essentially semiclassical in nature and involves just the classical BTZ black hole background. Moreover, we give arguments that the same mechanism for evading the black hole information paradox which we uncovered in the case of Ising CFT, should work essentially for arbitrary $CFT_2$ and thus also for $c\to \infty$, where a semiclassical picture is unambiguously expected to hold.
}

The outline of this paper is as follows. First we present the results of the Ising CFT calculation of the temporal evolution of a perturbation at nonzero temperature. Then we summarize, in a self-contained way, arguments for the existence of a holographic dual for Virasoro minimal model CFTs following from the Bershadsky-Ooguri construction. Subsequently we evaluate the thermal expectation value of the metric operator showing the appearance of the BTZ black hole in the high temperature limit, as well as a surprising fragility of the black hole at the quantum level. 
Section~\ref{s.main} comprises the main results of the paper, and can be read largely independently from sections~\ref{s.bo}-\ref{s.fragility}.
Here we describe the bulk Wilson line network interpretation of the Ising CFT results and the various phases of the evolution. 
We discuss the possible gravity interpretation of the observed recurrence in section~\ref{s.interpretation} \red{and comment on the generality of the proposed mechanism of evading the black hole information paradox in section~\ref{s.generality}.}
We conclude the paper with a summary and discussion and relegate some technical details to the Appendix.

\section{Ising CFT analysis}

We consider the CFT on a cylinder of size $R=2\pi$, which sets the scale for the temperature. The energy eigenvalues will then be exactly equal to the conformal dimensions $\Dl = h + \hb$.
In this paper, we focus on operators integrated over the spatial circle so as to be always in the $s$-channel and study just the temporal dependence.
\eq
O(t) =  \int_0^{2\pi} \f{dx}{2\pi} \OO{t,x}
\eqx
We will  perturb the thermal density matrix at $t=0$ by acting with $\exp(i \epsilon\, O(0))$ and consider just the term at linear order in $\epsilon$:
\eq
\rho \to e^{-i \epsilon\, O(0)} \rho e^{i \epsilon\, O(0)} \sim 
\rho + \epsilon \,\underbrace{i  [\rho,\, O(0)]}_{\dl\rho} + \ldots 
\eqx
Subsequently we evolve the perturbation $\dl\rho$ in time and measure the expectation value of the $O(t)$ operator
\eq
R_O(t) \equiv \cor{O(t)}_{\dl\rho(t)}\!\! = \tr O(t)\dl\rho(t) = 
\tr O(t) e^{-i H t} \dl\rho\, e^{i H t}
\eqx
This quantity can be numerically computed in a straightforward way using TCSA~\cite{Yurov:1989yu}. For the Ising CFT, however, we have at our disposal an exact expressions for two-point functions of the primary operators on the torus, from which we may extract the above quantity by integrating the \emph{retarded} two-point function at nonzero temperature along the spatial circle\footnote{More precisely, we would have two integrals -- for the source at $t=0$ and for the measurement at $t>0$, but this is equivalent due to translation invariance.}:
\eq
R_O(t) = -i\int_0^{2\pi} \f{dx}{2\pi} \th(t) \cor{\left[ \OO{t,x},\, \OO{0,0} \right]}
\eqx
Care must be taken in using the appropriate $i\epsilon$ prescription for the Wightman function $\GG$ being the analytical continuation of the Euclidean Green's function (see e.g. \cite{lorentzianCFT}). We obtain
\eqn
&&R_O(t) = \lim_{\epsilon \to 0^+} \int_0^{2\pi} \f{dx}{2\pi} \biggl[
\GG(x-t+i\epsilon, x+t-i\epsilon) - \nonumber\\ 
&&\hspace{3.5cm} \GG(x-t-i\epsilon, x+t+i\epsilon) \biggr]
\eqnx
where we also used the natural light-cone coordinates. 
In this paper, we will focus on the $\eps$ operator with $h=\hb=\f{1}{2}$, for which we can evaluate this expression by residues
\eq
\label{e.residues}
R_O(t) = \!\!\!\sum_{n=-\infty}^\infty \res_{x=t+2\pi n} \GG(x-t, x+t) -\!\!\!\!\!\!\! \res_{x=-t+2\pi n} \GG(x-t, x+t)
\eqx
where $x$ is constrained to lie in the interval $[0,2\pi)$.
The Euclidean two-point function on the torus can be found in \cite{DiFrancesco:1987ez,DiFrancesco:1997nk} and is given by
\eq
\label{e.torus}
\cor{\eps(z,\zb) \eps(0,0)} = \f{1/4}{Z_{ising}} \left| \f{\partial_z \th_1(0|\tau)}{\th_1(\f{z}{2}|\tau)} \right|^2 
\sum_{\nu=2}^4 \left| \f{\th_\nu(\f{z}{2}|\tau)^2}{2\eta(\tau) \th_\nu(0|\tau)} \right|
\eqx
where
\eq
Z_{ising} = \f{\th_2(0|\tau) + \th_3(0|\tau) + \th_4(0|\tau)}{2\eta(\tau)}
\eqx
The $1/4$ in (\ref{e.torus}) and the argument $\f{z}{2}$ follow from the \texttt{Mathematica} conventions for theta functions, which have spatial $\pi$ periodicity, while we consider a cylinder of size $2\pi$. 
In order to pass to Minkowski signature, we perform the substitutions
\eq
z = x-t  \qq \zb = x+t
\eqx
Evaluating the residues yields the final exact answer for the observable of interest
\eq
\label{e.Rexact}
R_\eps(t) = C \cdot \f{\th_2(t|\tau) + \th_3(t|\tau) +\th_4(t|\tau)}{\th_1(t|\tau)}
\eqx
with
\eq
C = \f{\th'_1(0|\tau)}{\th_2(0|\tau) + \th_3(0|\tau) + \th_4(0|\tau)} 
\eqx
and
\eq
\tau = \f{i}{2\pi T}
\eqx
In Fig.~\ref{fig.tcsa}, we show the time evolution of $R_\eps(t)$, compared to direct numerical TCSA computations\footnote{We use Ising CFT data from~\cite{Horvath:2022zwx} for setting up the TCSA.}, which just serve as an independent cross-check of the observed phenomena.

Expanding $R_\eps(t)$ in the large temperature limit, for $t<\pi/2$, we obtain
\eq
R_\eps(t) \sim \f{2\pi T}{1+ e^{-\f{\pi^2}{2} T}} \left[
\f{1}{\sinh 2\pi T t} + e^{-\f{\pi^2}{2} T} \coth 2\pi T t +\ldots \right]
\eqx
We observe two salient features -- first an exponential decay and then a plateau
\eq
\label{e.damping}
R_\eps(t) \sim 4\pi T \left[ e^{-2\pi T t} + \f{1}{2} e^{-\f{\pi^2}{2} T} +\ldots \right]
\eqx
The first term corresponds exactly to a quasi-normal mode (QNM) of a BTZ black hole\footnote{This is, in fact, an expected general result, see~\cite{Birmingham:2001pj,Son:2002sd}.}. 
The interpretation of the plateau is less clear. As the Ising CFT has a nonzero thermal average $\cor{\eps}_T \neq 0$ \cite{Gaberdiel:2008ma}, one could think that the plateau could come from a shift of the temperature due to the perturbation (``infalling matter''). 
The temperature shift would appear, however, only at a quadratic order in the perturbation as $\tr H \dl \rho(t)=0$.
We will address a black hole interpretation of the magnitude of the plateau later in the paper, corresponding to a loop diagram in the BTZ black hole (in fact quite similar to a scalar loop considered in~\cite{Kraus:2016nwo} for the torus 1-point function).

The final key feature of the exact formula~(\ref{e.Rexact}) is that it is periodic in time with periodicity $2\pi$ 
\eq
R_\eps(t+2\pi) = R_\eps(t) 
\eqx
More generally, the period would be equal to the size of the spatial circle $R$. From the CFT side, an approximate periodicity in $R$ is very easy to understand as the level spacing between the infinite number of descendants within a single representation is $2\pi/R$. Here, due to the conformal weights of the $\eps$ field and its fusion rules we have in fact exact periodicity.
In more generality, for any Virasoro minimal model CFT, we have always an exact periodicity (for any observable) with an appropriate mutliple of $R$, as there are just a finite number of conformal weights being rational numbers. For the Ising model indeed we have
\eq
e^{i 16 \pi H} = id
\eqx
The periodicity in time leads, on the other hand, to a crucial tension with a potential dual gravitational picture, \emph{if} there is in addition a black hole interpretation -- which is strongly suggested by the QNM-like initial exponential damping which we saw in~(\ref{e.damping}).

In the following part of the paper, we therefore review some arguments for the existence of a dual (quantum) bulk gravitational picture for the Virasoro minimal models and for the relevance of a black hole interpretation in the high temperature limit.
Then, in section~\ref{s.main}, we proceed to recover the three main features of the behaviour of $R_\eps(t)$ --- QNM damping, plateau and recurrence (periodicity) --- directly from the bulk perspective. We also comment on the ``burst'' seen in Fig.~\ref{fig.tcsa} at $t=\pi$.

\section{3D gravity and Chern-Simons theory}
\label{s.3dcs}

It is well established that 3D gravity with a negative cosmological constant can be recast as a pair of Chern-Simons theories with $SL(2,\RR)$ gauge groups~\cite{Achucarro:1986uwr,Witten:1988hc}. A general classical solution satsifying asymptotically AdS  boundary conditions (with our conventions) has the form:
\eqn
\label{e.aclassical}
A &=& b^{-1} \underbrace{\left( t^+ - l(z) t^{-} \right)}_{a(z)} b dz + b^{-1}\partial_\rho b d\rho\\
\label{e.abclassical}
\Ab &=& b \underbrace{\left( t^{-} - \lb(\zb) t^{+} \right)}_{\ab(\zb)} b^{-1} d\zb + b\partial_\rho b^{-1} d\rho
\eqnx
with $b\equiv b(\rho)$ being a gauge choice e.g. $b(\rho) = \exp (\rho\, t^0)$, 
\eq
z=x+it_E=x-t \qq \zb=x-it_E=x+t \ ,
\eqx
$t^\pm$ and $t^0$ are $SL(2,\RR)$ generators 
\eq
t^+ = \arr{0}{0}{-1}{0} \quad t^0=\arr{\f{1}{2}}{0}{0}{\red{-}\f{1}{2}} \quad
t^- = \arr{0}{1}{0}{0}
\eqx
satisfying
\eq
\label{e.sl2r}
[t^+, t^-] = 2 t^0 \qq [t^\pm, t^0] = \pm t^\pm
\eqx
while $l(z)$ and $\lb(\zb)$ are arbitrary holomorphic and antiholomorphic functions, which in the $AdS_3/CFT_2$ context are identified with $\f{6}{c} T(z)$ and $\f{6}{c} \Tb(\zb)$ (see below). The 3D metric is defined through
\eq
\label{e.metric}
g_{\mu\nu} = \f{1}{2} \tr\, (A_\mu-\Ab_\mu)(A_\nu-\Ab_\nu)
\eqx
and automatically solves 3D Einstein's equations.
For $l$ and $\lb$ being constant, one obtains a (possibly rotating) BTZ black hole~\cite{Banados:1992wn}, which reduces to the non-rotating case for $\lb=l$:
\eq
\label{e.btz}
ds^2 = -(e^\rho-l e^{-\rho})^2 dt^2 + (e^\rho+l e^{-\rho})^2 dx^2 + d\rho^2
\eqx
with $x$ being either noncompact or compact.
Note that the geometry, as given by the metric, and thus also the causal horizon structure is defined in terms of \emph{both} gauge fields $A$ and $\Ab$. E.g. fixing $l$ and changing $\lb$ would transform a non-rotating black hole into a rotating one and thus significantly change the horizon structure.
We will return to this point later in the paper.

If we formally insert the values of the parameters for the high temperature limit of the Ising CFT (see below for some justification), we get $l=\f{\pi^2}{12} T^2$. Then, the BTZ metric~(\ref{e.btz}) can be rewritten as (with $x \simeq x+2\pi$)
\eq
ds^2 = -(r^2-r_+^2) dt^2 + \f{dr^2}{r^2-r_+^2} + r^2 dx^2
\eqx
and $r_+ = 2\pi T$. For completeness, we quote the lowest QNM frequency for a massive scalar field with $m^2=\Dl (\Dl-2)$~\cite{Birmingham:2001pj}: 
\eq
\label{e.btzqnm}
\omega_{QNM}^{lowest}=-i \Dl r_+ = -i 2\pi T \Dl
\eqx
\red{Let us comment more on the BTZ black hole relevant for the present paper and its Chern-Simons description. It has a periodically identified spatial coordinate ($x \simeq x+2\pi$ as above), so the boundary at $\rho=\infty$ has the topology of a cylinder. The metric (\ref{e.btz}) corresponding to the Chern-Simons gauge fields (\ref{e.aclassical})-(\ref{e.abclassical}) with $b(\rho) = \exp (\rho\, t^0)$ is directly obtained in Fefferman-Graham (FG) coordinates, which normally extend only up to the horizon located at $e^{2\rho_*}=l$. The FG coordinate system breaks down at that point, but the Chern-Simons gauge fields are perfectly regular and extend beyond that up to $\rho=-\infty$, which is a singularity from the point of view of Chern-Simons fields (denoted by a red star in figures later in the paper), allowing for noncontractible Wilson loops, which will be essential in the following. This singularity is \emph{a-priori} distinct from the conventional causal black hole singularity which could be reachable in a variant of Eddington-Finklestein coordinates. However, the considerations of the present paper do not depend on these issues, apart just from the existence of noncontractible loops, so we will not explore this any further.}

\section{The Bershadsky-Ooguri construction}
\label{s.bo}

Even though the celebrated link between Wess-Zumino-Witten conformal field theories and Chern-Simons theories with the same gauge group~\cite{Witten:1988hf,Elitzur:1989nr} apparently could be thought of as an example of holography, it is in fact completely different in essence. E.g. the (usually compact) group $G$ of the CFT is the same as the gauge group of the CS. So there is no gravity in the bulk, which requires an $SL(2,\RR)$ CS, \emph{irrespective} of the CFT Kac-Moody symmetry group or even of its presence.

The above link can be formally used, however, directly for $G=SL(2,\RR)$, providing in principle\footnote{Due to the noncompact nature of the group, making this precise and rigorous on the quantum level may be quite nontrivial c.f.~\cite{Witten:2007kt,Cotler:2018zff,Porrati:2019knx}.} an equivalence between $SL(2,\RR)$ CS and an $SL(2,\RR)$ WZW. Imposing asymptotically $AdS$ boundary conditions involve setting the boundary values of the CS gauge fields to
\eq
A^+ \to 1 \qqqq \Ab^- \to 1 
\eqx
As shown in the seminal work~\cite{Coussaert:1995zp}, at the classical level one recovers the Virasoro algebra realized as asymptotic symmetries of the bulk spacetime.

As the boundary values of the CS gauge fields correspond to the current operators, the asymptotically $AdS$ boundary conditions mean in the $SL(2, \RR)$ Kac-Moody algebra that e.g. the $J^+$ field has to be set to 1.
In a remarkably prescient paper~\cite{Bershadsky:1989mf}, Bershadsky and Ooguri, showed that imposing\footnote{The paper considered explicitly $J^-=1$ which would be directly relevant for $\Ab$.} $J^+=1$ in the SL(2) Kac-Moody algebra through a BRST construction yields an isomorphism of BRST cohomology of the Kac-Moody representations (i.e. the remaining $SL(2, \RR)$ degrees of freedom after imposing the asymptotically $AdS$ boundary conditions) with minimal model Virasoro representations. 
It is important to note, that the above isomorphism was derived at a fully quantum level.

The non-vacuum Virasoro representations are mapped by~\cite{Bershadsky:1989mf} to the non-vacuum Kac-Moody representations, which in the dual Chern-Simons language (with $SL(2,\RR)$ gauge groups, so representing 3D gravity) are expected to correspond to an insertion of appropriate Wilson lines~\cite{Elitzur:1989nr} (see also~\cite{Verlinde:1989ua} for another perspective). Thus they correspond to matter fields\footnote{Here we differ from the proposed pure gravity (Euclidean) dual of the Ising model~\cite{Castro:2011zq}. Note however, that there exist diverse nonobvious equivalences between theories in 2-3 dimensions so there might not necessarily be a contradiction.} coupled to gravity.
A second quantized description of the appropriate matter fields is still missing and remains an outstanding problem (but see~\cite{Castro:2023bvo}).
The key result of this paper in fact comes from a close examination of the behaviour of the bulk matter fields with respect to gravity. 

An important ingredient of the Bershadsky-Ooguri construction is the twisting of the Kac-Moody energy-momentum tensor
\eq
\label{e.twisting}
T_{twisted} = T_{KM} + \partial J^0 
\eqx
which brings down the conformal weight of $J^+$ to 0, allowing it to be consistently set to a constant, raises $J^-$ to~2 and keeps $J^0$ at 1.
As we will see below, this has clear implications from the point of view of the (quantized) bulk Chern-Simons theory. 

\section{Chern-Simons gauge fields as quantum operators in the CFT}

Quantizing the $SL(2, \RR)$ Chern-Simons theory remains challenging (c.f.~\cite{Witten:2007kt,Cotler:2018zff,Porrati:2019knx}), so we will adopt here a very heuristic, holography inspired approach.
We may start from a general classical solution of the Chern-Simons theory
\eq
\label{e.csclassical}
A(z, \zb, \rho) = b(\rho)^{-1}a(z)b(\rho)\,dz + b^{-1}\partial_\rho b\, d\rho
\eqx
where the gauge transformation is typically chosen to be $b(\rho) = \exp (\rho\, t^0)$ for a geometrical interpretation with $\rho=+\infty$ being the boundary, or $b(\rho)=1$ for ease of computation with Wilson line networks and 
\eq
a(z) = a^+(z) t^+ + a^0(z) t^0 + a^-(z) t^- 
\eqx
For quantization, we would like to promote the remaining parameters/functions appearing the general classical solution~(\ref{e.csclassical}) to operators --- here the $a^\pm(z)$ and $a^0(z)$ functions. From the perspective of holography, these should be operators from the minimal model CFT and the twisting in the Bershadsky-Ooguri construction essentially fixes these operators uniquely.
Indeed, as the components of the remaining field $a(z)$ can be understood as being the currents of the Kac-Moody algebra (see~\cite{Banados:1994tn,Banados:1998gg}), we should set $a^+(z)=1$ (as $J^+$ has dimension 0 after the twisting~(\ref{e.twisting}), it should be proportional to the identity operator). Since $J^0(z)$ has dimension 1 and such a field is absent in the vacuum Virasoro representation of a CFT, we set $a^0(z)=0$. $J^-(z)$ has dimension 2, hence it should be proportional to the \emph{unique} field with dimension 2 in the vacuum representation --- the energy-momentum tensor $T(z)$. Hence we obtain
\eq
a(z) = t^+ + \alpha T(z) t^-
\eqx
This realizes the (quantum) CS gauge fields as \emph{concrete operators} in the boundary minimal model CFT. 
The coefficient $\alpha$ can be fixed to be $\alpha=6/c$ so that evaluating for the CFT vacuum state reproduces the global $AdS_3$ spacetime\footnote{It would be interesting, however, to derive the value of $\alpha$ directly on the quantum level.}.
This yields finally
\eqn
\label{e.quantumA}
\hspace{-0.75cm}\widehat{A}(z, \zb, \rho)\!\! &=&\!\! b^{-1} \left( t^+ - \f{6}{c}T(z) t^{-} \right) b dz + b^{-1}\partial_\rho b d\rho \\
\label{e.quantumAb}
\hspace{-0.75cm}\widehat{\Ab}(z, \zb, \rho)\!\! &=&\!\!  b \left( t^{-}\!\!\! - \f{6}{c}\Tb(\zb) t^{+} \right) b^{-1} d\zb + b\partial_\rho b^{-1} d\rho
\eqnx
which look like the classical answers given previously in (\ref{e.aclassical})-(\ref{e.abclassical}), but are now considered as quantum operators in the minimal model CFT, with $T(z)$ and $\Tb(\zb)$ being the quantum energy-momentum tensors. 
Of course, these formulas are not new and have been studied in~\cite{Bhatta:2016hpz,Besken:2016ooo,Fitzpatrick:2016mtp,Besken:2017fsj,Hikida:2017ehf,Hikida:2018dxe,Besken:2018zro} for generic CFTs. Evaluating the thermal average of these operators recovers trivially the expressions for the classical BTZ black hole 
\eq
\label{e.Athermal}
\cor{\widehat{A}(z,\zb,\rho)}_T\!\!\! = A_{BTZ}(\rho)  \quad \cor{\widehat{\Ab}(z,\zb,\rho)}_T\!\!\! = \Ab_{BTZ}(\rho)
\eqx
even for the Ising CFT, thus supporting the expectation that we are dealing with a (potentially quantum) black hole on the gravitational side.

\section{Fragility of quantum black holes}
\label{s.fragility}

It is interesting to use formulas (\ref{e.quantumA}-\ref{e.quantumAb}) to explicitly present the bulk metric as a quantum operator in the boundary CFT:
\eqn
\label{e.quantummetric}
\widehat{g_{\mu\nu}}dx^\mu dx^\nu &=& \f{6}{c}T(z)\, dz^2 + \f{6}{c}\Tb(\zb)\, d\zb^2 + \nonumber\\
&& \hspace{-1.5cm}  +\left( e^{2\rho} + \f{36}{c^2} T(z) \Tb(\zb) e^{-2\rho} \right)dz d\zb +d\rho^2
\eqnx
Such a formula (understood as a quantum operator) appeared first in~\cite{Banados:1998gg}, but due to a lack of holographic interpretation it was not clear from which CFT the energy-momentum tensor is taken. We will now compute the expectation value of this operator.

Let us note that the expectation value of (\ref{e.quantummetric}) is a solution of 3D Einstein's equations only if $T(z)\Tb(\zb)$ would factorize in the expectation value
\eq
\cor{T(z)\Tb(\zb)} \simeq \cor{T(z)} \cor{\Tb(\zb)}
\eqx
It is then a (nonrotating) BTZ black hole solution for $\f{6}{c}\cor{T(z)}=\f{6}{c}\cor{\Tb(\zb)}=l>0$:
\eq
\label{e.btzii}
ds^2_{BTZ} = -(e^\rho-l e^{-\rho})^2 dt^2 + (e^\rho+l e^{-\rho})^2 dx^2 + d\rho^2
\eqx
This happens for the Ising CFT in the limit of high temperatures thus supporting the interpretation of dealing with a black hole on the dual side\footnote{Of course, this may be far from a classical geometry {\it per se} as we are only making a statement for the expectation value of the metric operator.}. For finite temperatures, the situation is, however, surprisingly more subtle.

\begin{figure*}[t]
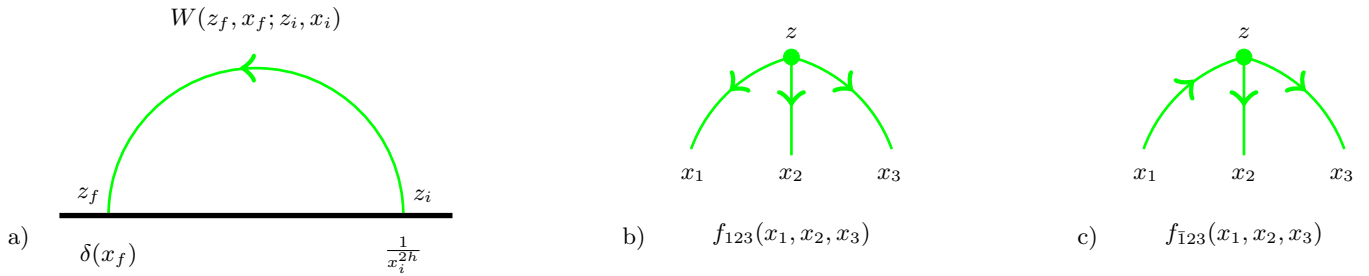

    \centering
    \figrules
    \caption{a) Wilson line, together with outgoing and ingoing external states on the boundary. b) and c) The bulk vertex with two different orientations for the first Wilson line. The bulk vertices do not depend on $z$. Only the holomorphic Wilson lines/external states are shown.}
    \label{fig.rules}
\end{figure*}

Zamolodchikov showed~\cite{Zamolodchikov:2004ce} that the product $T(z)\Tb(\zb)$ does not have any divergences or ordering ambiguities and that for CFT states with \emph{definite energy and momentum} the expectation value indeed factorizes
\eq
\cor{T(z)\Tb(\zb)} = \cor{T(z)} \cor{\Tb(\zb)}
\eqx
However, it is easy to see that this factorization breaks down already if the state is a linear combination of two basis states from two \emph{different} Virasoro representations\footnote{If the states were from the same Virasoro representation, one could have a nontrivial matrix element of $L_n$ between them and thus, generically, a more complicated $z$-dependent metric.}. Indeed, taking for simplicity
\eq
\ket{\Psi} = \sqrt{1-p} \ket{0} + \sqrt{p} \ket{E}
\eqx
with $\cor{T(z)}_0=0$ and $\cor{T(z)}_E=E$ and similarly for the antiholomorphic ones, we get
\eq
\cor{T(z)\Tb(\zb)}_\Psi = p E^2 \geq p^2 E^2 = \cor{T(z)}_\Psi \cor{\Tb(\zb)}_\Psi
\eqx
with equality only if $p=0$ or $p=1$. In Appendix~\ref{s.inequality}, we discuss some more general conditions when this inequality holds. 
Defining
\eq
r = \cor{T(z)\Tb(\zb)}_T - \cor{T(z)}_T \cor{\Tb(\zb)}_T \ ,
\eqx
we checked numerically that $r>0$ for the thermal density matrix of the Ising CFT, and vanishes in the high temperature limit.
The above result implies in particular that the thermal average geometry has a nontrivial correction w.r.t. the classical BTZ black hole. In Euclidean signature we thus have
\eq
\cor{\widehat{ds^2}}_T = ds^2_{BTZ} + \f{36}{c^2} r \cdot  e^{-2\rho} (dt_E^2 +dx^2)
\eqx
Due to the positive value of $r>0$, the Euclidean geometry is no longer capped at the horizon but a throat opens up.


This result indicates that quantum black holes seem to be very fragile w.r.t. quantum effects as any admixture of a state from another Virasoro representation manifests itself in the appearance of a throat. That being said, we note, however, that it is not clear what is the impact of the \emph{average} geometry on the quantum dynamics of matter. We leave this for further investigation.
In particular, the surprising (from the black hole perspective) recurrence/periodicity observed in the Ising model is \emph{not related} to this phenomenon at all.
Indeed, it exists in the same form for any value of the temperature, while the throat coefficient $r$ strongly depends on temperature.

\section{Bulk Wilson line network interpretation of Ising evolution}
\label{s.main}

\begin{figure*}[t]
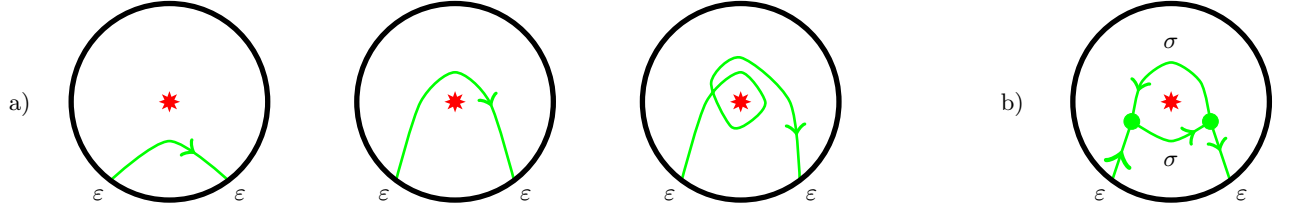

    \centering
    \raisebox{1.3cm}{a)}\hspace{0.5cm}\qnma
    \hspace{1cm}\qnmb\hspace{1cm}\qnmc
    \hspace{2cm}\raisebox{1.3cm}{b)}\hspace{0.5cm}
    \plateau
    \caption{a) Sample Wilson lines appearing in the QNM regime calculation. The high temperature answer is dominated by the first graph. b) Leading graph contributing to the plateau. Only the holomorphic Wilson lines/external states are shown.}
    \label{fig.qnm}
\end{figure*}

We will now proceed to the main part of the paper and analyze how the key features of the various stages of the CFT temporal evolution arise from the bulk gravitational perspective. The key question of interest is the bulk mechanism behind the recurrence i.e. the re-emergence of the signal from the black hole.

\red{Recall that in the Chern-Simons description, gravity is represented by the pair of Chern-Simons gauge fields $A^a_\mu$ and $\Ab^a_\mu$, with the bulk metric (and thus the causal structure) being given by (\ref{e.metric}). 
Their thermal expectation values~(\ref{e.Athermal}) represent the BTZ black hole solution~(\ref{e.btz}).
The CFT primary fields, on the other hand, correspond to bulk matter fields, which are realized through appropriate Wilson lines as we describe below.}

CFT correlation functions can be computed on the gravitational Chern-Simons side through a network of Wilson lines. This approach has been developed in~\cite{Bhatta:2016hpz,Besken:2016ooo,Fitzpatrick:2016mtp,Besken:2017fsj,Hikida:2017ehf,Hikida:2018dxe,Besken:2018zro}. Here we follow the formulation from~\cite{Fitzpatrick:2016mtp} with some minor modifications\footnote{In particular we change the sign of $T(z)$ so as to have $Re\, z$ as a spatial coordinate. We also rescale $L^-$ for consistency.}. 
The Wilson lines extend to the boundary to the insertion points of the operators appearing in the CFT correlation function of interest 
\eq
\label{e.wilsongen}
P e^{\int A^a_\mu L^a dx^\mu}
\eqx
and the $L^a$ are $SL(2,\RR)$ generators in an appropriate $SL(2,\RR)$ representation characteristic of the conformal weight. 
It is convenient to use, following~\cite{Fitzpatrick:2016mtp}, a realization of the generators in terms of differential operators in \red{an auxiliary variable} $x$:
\eqn
\label{e.diffgen}
L^+ &=& \partial_x \nonumber\\
L^0 &=& x\partial_x + h \nonumber\\
L^- &=& x^2 \partial_x +2 h x
\eqnx
satisfying the commutation relations (\ref{e.sl2r}).
Note that we leave a concrete specification of the Hilbert space of the representation for future investigation, as it is not completely clear what would be the best formulation -- whether one should use individual $SL(2,\RR)$ representations or rather a joint $SL(2,\CC)$ one or adopt some specific contour choices for the $x$ integration etc. Our considerations in this paper, however, do not depend on these choices. \red{We caution the reader, that the auxiliary variable $x$ should not be confused with the spatial boundary coordinate of the Ising model CFT. Indeed, in all our computations using the auxiliary variable $x$ we will use holomorphic notation $z$ and $\zb$ for the CFT coordinates, and only transition to $(x, t)$ in the final formulas.}

The Wilson lines may merge in the bulk in triple vertices whose specific form follows from requiring $SL(2,\RR)$ gauge invariance and is, in addition, multiplied by the relevant OPE coefficient. The Wilson lines are present only for the matter fields of the theory. 
One can think of the Wilson line networks as providing a first quantized formulation of (topological) bulk matter fields dual to the nontrivial CFT primary operators.

The Wilson line network is gauge invariant and is evaluated in the topological Chern-Simons theory, hence the contours can be continuously deformed in an arbitrary way. Hence it is standard to simplify the final explicit calculation by passing to the (non-geometric) simple gauge $b(\rho)=1$ so that
\eq
\label{e.A}
A = \left(L^+ - \f{6}{c} T(z) L^{-} \right) dz 
\eqx
and to effectively collapse the curves on to the boundary. Naturally, one is guaranteed to obtain the same answer when evaluating the Wilson line network in the geometric gauge $b(\rho) = \exp (\rho\, t^0)$. 
For the antiholomorphic sector we have
\eq
\label{e.Ab}
\Ab = \left(L^- - \f{6}{c} \Tb(\zb) L^{+} \right) d\zb
\eqx
and we comment on the appropriate modifications in Appendix~\ref{s.antiholo}. But clearly the final antiholomorphic results have to be analogous to the holomorphic ones with the modification $z\to\zb$ and $h\to \hb$.

The last ingredients are the $\ket{h}$ and $\bra{h}$ states in the $x$-representation which have to be inserted at the boundary endpoint of the Wilson line depending on its orientation. 
\eq
\ket{h} = \f{1}{x^{2h}} \qqqq \bra{h} = \dl(x)
\eqx
\red{satisfying $L^-\ket{h}=0$ and $L^0\ket{h}=-h\ket{h}$.}
We summarize these rules in Fig.~\ref{fig.rules}. The purely outgoing triple vertex is given by
\eq
f_{123}(x_1,x_2,x_3) = \f{1}{x_{12}^{h_1+h_2-h_3} x_{23}^{h_2+h_3-h_1} x_{31}^{h_3+h_1-h_2}}
\eqx
while changing a line to ingoing involves changing $h_i \to 1-h_i$ and adding a nontrivial normalization\footnote{Which can presumably be fixed using formulas from~\cite{Belavin:2024nnw}.}. However, we will not use these expressions directly in explicit computations in the present paper.

\red{The Wilson line matrix element $W(z_f, x_f; z_i, x_i)$ can be evaluated in a very convenient form using a Hamiltonian approach~\cite{Fitzpatrick:2016mtp}, which we briefly summarize below. The path ordered exponential
\eq
P \exp \left( \int_{z_i}^{z_f} A(z) \right)
\eqx
can be interpreted as a time evolution (with the holomorphic coordinate $z$ playing the role of time) of a time-dependent Hamiltonian $H=i A(z)$ being an operator in the auxiliary $x$-space. The derivative $\partial_x$ can be rewritten in terms of the conjugate momentum $p=-i\partial_x$. Subsequently, with the above identifications the evolution operator $W(z_f, x_f; z_i, x_i)$ may be written in a standard way in a path integral form
\eq
\int \DD p(z) \int_{x(z_i)=x_i}^{x(z_f)=x_f} \DD x(z) e^{\int_{z_i}^{z_f} \left[ ip\left(\f{dx}{dz} +1 -\f{6T}{c} x^2\right) - \f{6T}{c} 2h x \right] dz }
\eqx
using (\ref{e.A}) and the explicit form of the $SL(2,\RR)$ generators (\ref{e.diffgen}).
Integrating out $p(z)$ leads to a differential equation for $x(z)$ and we obtain a simple closed form formula:
}
\eq
\label{e.wa}
W(z_f, x_f; z_i, x_i) = e^{-\int_{z_i}^{z_f} dz \f{12h T(z)}{c} x_T(z)} \cdot 
\dl(x_i - x_T(z_i))
\eqx
where $x_T(z)$ is the solution of
\eq
\label{e.wb}
-x'(z) = 1 - \f{6 T(z)}{c} x^2(z)
\eqx
with the boundary condition
\eq
\label{e.wc}
x_T(z_f) = x_f
\eqx
In general the above formulas are treated as quantum operators through the presence of the energy-momentum tensor. Ref.~\cite{Fitzpatrick:2016mtp} introduced a normal ordering prescription which allows for contractions of the energy-momentum tensors only between distinct lines.

The factor of $1/c$ accompanying the energy-momentum tensor ensures a regime of semiclassical gravity by suppressing contractions of the energy-momentum tensor in the $c \to \infty$ limit. 
At large temperatures instead, we use the thermal expectation values of $T(z)$ in these formulas
\eq
\cor{ e^{\ldots T(z) } \cdot e^{\ldots T(z') }\cdot \ldots}_T
\longrightarrow
e^{\ldots \cor{T(z)}_T } \cdot e^{\ldots \cor{T(z')}_T }\cdot \ldots
\eqx
and thus the contractions would have to be done using a thermal \emph{connected} correlation function
\eq
\label{e.TTconn}
\cor{T(z) T(z')}_T - \cor{T(z) }_T \cor{T(z')}_T
\eqx
which is largely suppressed at high temperatures.
Indeed, the connected correlation function (\ref{e.TTconn}) is known exactly for the Ising model~\cite{DiFrancesco:1987ez}. The high temperature limit of the expressions given in Appendix~\ref{s.tt} yields
\eq
\label{e.TTconnhighT}
\f{1}{4} \f{\pi^4 T^4}{1+e^{-\f{\pi^2}{2} T}} \left[ 
\f{1}{\sinh^4 \pi T z} + e^{-\f{\pi^2}{2} T} \f{\cosh 2\pi T z}{\sinh^4 \pi T z}
\right] + \ldots
\eqx
where $z$ is the difference of the arguments in (\ref{e.TTconn}) and is assumed to be real and less than $\pi$. We see that for $z$ of order 1, the connected correlator is indeed exponentially suppressed. Care must be taken, however, for small arguments $z \ll 1/(\pi T)$, as then clearly the UV singularity $(c/2)/z^4$ prevails and these contractions are not suppressed, so not everything can be easily computed.

Let us now evaluate the Wilson line $W(z_f, x_f; z_i, x_i)$ for the case of constant $T(z)$ equal to the thermal expectation value which we parametrize by
\eq
\cor{T(z) }_T = \f{c}{6} a^2 
\eqx
with
\eq
a = \pi T
\eqx
in the high temperature limit.
Then the solution of (\ref{e.wb}) is
\eq
x_T(z_i) = \f{a x_f \cosh a z_{fi}+ \sinh a z_{fi}}{a(\cosh a z_{fi}+ a x_f\sinh a z_{fi})}
\eqx
and the Wilson line $W(z_f, x_f; z_i, x_i)$ can be evaluated to
\eq
\label{e.wilsongen}
W = \left( \cosh a z_{fi} + a x_f \sinh a z_{fi} \right)^{-2h} 
\dl\left( x_i - x_T(z_i) \right)
\eqx

\subsection{QNM regime}

Let us now compute the CFT 2-point function at nonzero temperature. 
As a warm-up, we first consider the infinite volume case.
According to the rules in Fig.~\ref{fig.rules}, we have to compute
\eq
\label{e.wtwop}
\int \dl(x_f) W(z_f, x_f; z_i, x_i) \f{1}{x_i^{2h}} dx_i dx_f
\eqx
Since $x_f=0$, the computation simplifies, so it is instructive to perform it directly.
The solution of (\ref{e.wb})-(\ref{e.wc}) is
\eq
x_T(z) = -\f{1}{a} \tanh a(z-z_f)
\eqx
Let us evaluate first the integral in~(\ref{e.wa})
\eq
e^{-2h a^2 \int_{z_i}^{z_f} x_T(z) dz} = e^{2h a \int_{z_i}^{z_f} \tanh a(z-z_f) dz} 
= \f{1}{(\cosh a z_{fi})^{2h}}
\eqx
At large temperatures, the $tanh$ is a constant so if $z_i$ and $z_f$ are not close we get exponential damping. Plugging the above into (\ref{e.wtwop}) we get finally
\eq
\left[ \f{1}{\cosh a z_{fi}} \cdot \f{a}{\tanh a z_{fi}} \right]^{2h}
= \left( \f{a}{\sinh a z_{fi}} \right)^{2h}
\eqx
Combining with an analogous antiholomorphic contribution yields
\eq
\label{e.ginfinite}
 \left( \f{a}{\sinh a z_{fi}} \right)^{2h}  \left( \f{a}{\sinh a \zb_{fi}} \right)^{2\hb}
\eqx
which is the exact thermal 2-point function in \emph{infinite} volume.

Let us now generalize to the finite volume case. Firstly, we will have to sum over all topologically inequivalent Wilson line networks. So, in particular, we will consider lines from $z_i$ to $z_f$ which will wrap the cylinder $n\in \ZZ$ times. These lines are inequivalent as the CS solution is defined on the punctured disk (times time). Ignoring any possible energy-momentum tensor contractions\footnote{These should be in principle forbidden due to the normal ordering prescription of~\cite{Fitzpatrick:2016mtp}, however the finite temperature version may require some care.}, yields
\eq
\label{e.mirror}
\sum_{n=-\infty}^\infty \left( \f{a}{\sinh a (z_{fi} + 2 \pi n)} \right)^{2h}
\eqx
which is just equivalent to summing over mirror images (here we ignore for the moment the antiholomorphic contributions, which, in fact, will be very important later).
For $2\pi \gg z_{fi} \gg \f{1}{a}$, which is the QNM regime, we can neglect all mirror images, reduce to $n=0$, and after incorporating the antiholomorphic part recover the same result~(\ref{e.ginfinite}) as in the infinite volume case.
In order to pass to our observable we specialize to $\eps$ with $h=\hb=\f{1}{2}$ and compute the residues
\eq
\label{e.residuesexample}
\left(\res_{x=t} - \res_{x=-t}\right) \left( \f{a}{\sinh a (x-t)} \right)  \left( \f{a}{\sinh a (x+t)} \right)
\eqx
to obtain
\eq
R_\eps(t) \sim \f{2 \pi T}{\sinh 2\pi T t} \sim 4\pi T\, e^{-2\pi T t}
\eqx
This reproduces the leading Ising model behaviour in~(\ref{e.damping}) and agrees with the BTZ QNM for $\Dl=1$ (see~(\ref{e.btzqnm}). These results are essentially fixed by symmetries, but the above computation shows how they follow explicitly in the Wilson line language.
We will now move to nonuniversal phenomena, in order to investigate in further detail the link with the bulk theory. 


\begin{figure}[t]
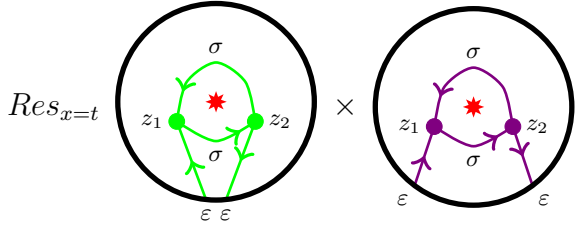

    \centering
    \raisebox{1.5cm}{\large $ \res_{x=t}$}\hspace{0.2cm}\plateaubigb
    \hspace{0.1cm}\raisebox{1.5cm}{\large $ \times$}\hspace{0.2cm}\plateaubig
    \caption{In order to compute the behaviour of $R_\eps(t)$ in the plateau region, one needs to compute the residue in one sector, and a graph with well separated legs in the other. In this paper, we just estimate the magnitude of the $\sg$-loop.}
    \label{fig.plateau}
\end{figure}

\subsection{The plateau --- ``quantum halo''}

Apart from the winding Wilson lines, for the Ising model we have nontrivial fusion $\sg \times \sg \to \eps$, and hence we have additional graphs with the $\eps$ Wilson line splitting into two $\sg$ Wilson lines which then wrap around the singularity with some net winding. A sample graph is shown in Fig.~\ref{fig.qnm}b. 

In order to compute the observable $R_\eps(t)$, one has to combine the holomorphic and antiholomorphic pieces and compute residues as in~(\ref{e.residuesexample}) (see Fig.~\ref{fig.plateau}). For that, however, the insertion points in one of the two sectors would have to be very close together, hence contractions through the connected thermal energy-momentum correlator would no longer be negligible (see discussion below~(\ref{e.TTconnhighT})). Such contractions would be allowed as they would extend between different Wilson line segments.
We leave this computation as a nontrivial but important open problem. Here we concentrate just on estimating the overall magnitude of the answer, which will most probably be governed by the two Wilson lines encircling the singularity.

The lower line is
\eq
\left( \cosh a z_{21} + a x_2 \sinh a z_{21} \right)^{-2h_\sg} 
\dl\left( x'_1 - x_T(z_1) \right)
\eqx
while the upper one is
\eq
\left( \cosh a \widetilde{z_{21}} + a x_1 \sinh a \widetilde{z_{21}} \right)^{-2h_\sg} 
\dl\left( x'_2 - x_T(z_2) \right)
\eqx
where $\widetilde{z_{21}} = 2\pi-z_{21}$. Then, when\footnote{In fact, both inequalities are not strictly necessary.} $a z_{21}\gg 0$ and $a(2\pi-z_{21})\gg 0$, the product scales like $e^{- 2\pi a \cdot  2 h_\sg}$. Putting together a similar factor from the antiholomorphic part, we obtain the estimate
\eq
e^{- 4\pi a \cdot  2 h_\sg} = e^{-\f{1}{2} \pi^2 T}
\eqx
which agrees with the magnitude of the plateau in the Ising CFT results~(\ref{e.damping}).
Perhaps not unexpectedly, the above result also agrees with a scalar loop around the horizon of the BTZ black hole, which was used in evaluating torus 1-point functions in~\cite{Kraus:2016nwo}:
\eq
e^{-2\pi \Dl r_+} = e^{-\f{1}{2} \pi^2 T}
\eqx
with $\Dl=\f{1}{8}$ being the dimension of the $\sg$ field.
Note, however, that we will observe a significant departure from the behaviour of scalar fields when we discuss the re-emergence of the signal in section~\ref{s.recurrence}.\\

\subsection{2-point a-cycle torus blocks}

Before we proceed, it is illuminating to compare the preceding ``bulk'' Chern-Simons expressions with 2-point conformal blocks on a torus. The standard ($s$-channel) 2-point blocks are typically denoted by
\[
\twop{$\varepsilon$}{$\varepsilon$}{$\varepsilon$}{$1$}{}
\]
where the loop represents the {\bf b}-cycle of the torus\red{, i.e. the compactified Euclidean time circle.}
For direct comparison with the Chern-Simons computation, one should, however, use a variant with the loop being associated to the {\bf a}-cycle instead\red{ --- the spatial circle}. In Appendix~\ref{s.torusblocks}, we give the relation to the standard ones.
It is convenient to introduce the normalized blocks\footnote{Recall that we are using Mathematica notation and $z$ with $2\pi$ periodicity.}
\eqn
f^1_{\eps\eps\eps}(z) = \f{1}{Z_{ising}}\!\! \twop{$\varepsilon$}{$\varepsilon$}{$\varepsilon$}{$1$}{\textbf{a}}\!\!\!\!\!\! &=&\!\!
\widetilde{C}\, \f{\th'_1(0)}{\th_1(\f{z}{2})} 
\left[ \f{\th_3(\f{z}{2})}{\sqrt{\th_3(0)}} + \f{\th_2(\f{z}{2})}{\sqrt{\th_2(0)}}\right] \nonumber\\
f^\eps_{\eps 1\eps}(z) = \f{1}{Z_{ising}}\!\! \twop{$\varepsilon$}{$\varepsilon$}{$1$}{$\varepsilon$}{\textbf{a}}\!\!\!\!\!\! &=&\!\!
\widetilde{C}\, \f{\th'_1(0)}{\th_1(\f{z}{2})} 
\left[ \f{\th_3(\f{z}{2})}{\sqrt{\th_3(0)}} - \f{\th_2(\f{z}{2})}{\sqrt{\th_2(0)}}\right] \nonumber\\
f^\sg_{\eps\sg\eps}(z) = \f{1}{Z_{ising}}\!\! \twop{$\varepsilon$}{$\varepsilon$}{$\sg$}{$\sg$}{\textbf{a}}\!\!\!\!\!\! &=&\!\!
\widetilde{C}\, \f{\th'_1(0)}{\th_1(\f{z}{2})} 
 \f{\sqrt{2}\th_4(\f{z}{2})}{\sqrt{\th_4(0)}} 
 \label{e.blocks}
\eqnx
where $0$ and $z$ are the insertion positions of the $\eps$ operators and
\eq
\widetilde{C} = \f{1}{2}\f{1}{Z_{ising}} \f{1}{2\sqrt{\eta(\tau)}}
\eqx
These blocks also include the OPE coefficients of the theory. 
It is clear that $f^\sg_{\eps\sg\eps}(z)$ directly resembles the graph in Fig.~\ref{fig.qnm}(b), while $f^1_{\eps\eps\eps}(z)$ and $f^\eps_{\eps 1\eps}(z)$, resemble the first two graphs in Fig.~\ref{fig.qnm}(a). The plateau value indeed follows  from the high temperature asymptotics of $f^\sg_{\eps\sg\eps}(z)$.
The Wilson lines wrapping the singularity multiple number of times are included in the 2-point blocks in a specific way.
Indeed, the normalized {\bf a}-cycle torus blocks have high temperature asymptotics of the form
\eqn
\label{e.even}
f^1_{\eps\eps\eps}(z) &\sim& \sum_{n=-\infty}^\infty \f{\pi T}{\sinh \left(\pi T (z + 2\pi \cdot 2n)\right)} \\
f^\eps_{\eps 1\eps}(z) &\sim& \sum_{n=-\infty}^\infty \f{-\pi T}{\sinh \left(\pi T (z + 2\pi \cdot (2n+1))\right)}
\label{e.odd}
\eqnx
so even and odd wrappings fall respectively into $f^1_{\eps\eps\eps}(z)$ and $f^\eps_{\eps 1\eps}(z)$. 
Note that all these terms correspond to the Chern-Simons expressions appearing in~(\ref{e.mirror}). 
The distinction of even and odd wrapping is consistent with the relation
\eq
\label{e.blockperiodicity}
f^1_{\eps\eps\eps}(z +2\pi) = - f^\eps_{\eps 1\eps}(z)
\eqx
In terms of the normalized torus \textbf{a}-cycle blocks, the 2-point function on the torus $\cor{\eps(z,\zb) \eps(0,0)}$ is given simply by
\eq
\label{e.eetorusblocks}
f^1_{\eps\eps\eps}(z) f^1_{\eps\eps\eps}(\zb) +
f^\eps_{\eps 1\eps}(z) f^\eps_{\eps 1\eps}(\zb) +
f^\sg_{\eps\sg\eps}(z) f^\sg_{\eps\sg\eps}(\zb)
\eqx
similarly as for the standard torus \textbf{b}-cycle blocks as in~\cite{DiFrancesco:1997nk}.
These observations will be very important for understanding the phenomenon of recurrence/periodicity, as discussed in the following section.

\subsection{The recurrence and its implications for gravity}
\label{s.recurrence}

\begin{figure}[t]
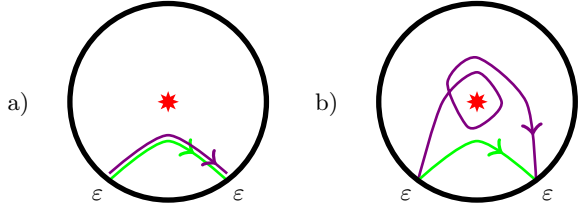

    \centering
    \raisebox{1.3cm}{a)}\hspace{0.5cm}\mirrora
    \hspace{0.5cm}\raisebox{1.3cm}{b)}\hspace{0.5cm}\mirrorb
    \caption{Two options for combining holomorphic and antiholomorphic Wilson lines in the nontrivial BTZ topology: a) coordinated windings as for a scalar field, b) largely independent windings as dictated by the boundary CFT.}
    \label{fig.mirror}
\end{figure}

Let us now return to the key issue of recurrence/periodicity --- the recovery of the whole signal after time $t=2\pi$, despite the fact that earlier it seemed as if the whole signal has been completely absorbed by the black hole through QNM exponential decay. Note that the physics of the plateau is really irrelevant for this question, as it is exponentially suppressed at large temperatures, while the recurrence occurs irrespective of the value of the temperature.

Recall the formula~(\ref{e.mirror}) for the Wilson line network summed over the winding around the singularity or equivalently summed over the mirror images:
\eq
\sum_{n=-\infty}^\infty \left( \f{a}{\sinh a (z_{fi} + 2 \pi n)} \right)^{2h}
\eqx
Note, however, that this expression only involves the holomorphic part of the answer. As $\eps$ has conformal weights $h=\hb=\f{1}{2}$, we certainly have to incorporate here also the nonholomorphic part.

As we can intuitively consider the Wilson line network as coming from some kind of (``topological'') path integral over trajectories of a particle coupled to gravity in the guise of the Chern-Simons gauge potentials $A$ and $\Ab$, it is natural to expect that the Wilson lines will be \emph{simultaneously} charged under both $A$ and $\Ab$ with weights $h=\hb=\f{1}{2}$ (as in Fig.~\ref{fig.mirror}a). This would lead to the following formula, specialized to these conformal weights
\eq
\label{e.eqmirror}
\sum_{n=-\infty}^\infty  \f{a}{\sinh a (z_{fi} + 2 \pi n)}  \cdot
\f{a}{\sinh a (\zb_{fi} + 2 \pi n)} 
\eqx
with $z=x-t$ and $\zb=x+t$.
The above formula, however, does not lead to a recurrence in the dynamics. 

Indeed, let us consider $t=2\pi + s$, with $s$ positive and small\footnote{This assumption is made for simplicity, just to ensure a unique choice of windings when computing residues.}.
In order for the poles for $x$ to be in the interval $[0, 2\pi)$, we have to choose $x=t - 2\pi=s$ ($n=1$) and $x=-t+4\pi=2\pi-s$ ($n=-2$). Computing the residues as in (\ref{e.residues}) yields
\eq
\f{2a}{\sinh 2a (2\pi+s)}
\eqx
which is exponentially suppressed.
This is in marked contrast with the periodic outcome observed in the CFT which in the high temperature limit and for small $s$ is instead
\eq
\label{e.periodic}
\f{2a}{\sinh 2 a s}
\eqx
Indeed, (\ref{e.eqmirror}) is the result that one would get by computing the scalar 2-point function in a finite volume BTZ background from the inifnite volume one through the method of images.
Clearly, a scalar perturbation in a black hole background is expected to be absorbed and would never re-emerge \emph{completely intact}.

In fact, as for the Ising model CFT we know that the relevant parts of the thermal 2-point functions are products of the torus \textbf{a}-cycle blocks being ratios of $\th$ functions (see (\ref{e.eetorusblocks}) and (\ref{e.even})-(\ref{e.odd})), the Ising CFT result in the large temperature limit is not (\ref{e.eqmirror}) but rather
\eq
\label{e.isingwindings}
\left( \sum_{\substack{n,n'\\ even}} + \sum_{\substack{n,n'\\ odd}} \right) \left[ 
\f{a}{\sinh a (z_{fi} + 2 \pi n)}  \cdot
\f{a}{\sinh a (\zb_{fi} + 2 \pi n')}
\right]
\eqx

Let us now repeat the residue computation. The pole at $x=t - 2\pi=s$ corresponds to $n=1$, hence the residue will involve a sum over all \emph{odd} $n'$. The dominant term for small $s$ will be the one with $n'=-1$:
\eq
\f{a}{\sinh 2a s}
\eqx
The pole at $x=-t+4\pi=2\pi-s$ corresponds to $n'=-2$, hence the residue will involve a sum over all \emph{even} $n$. The dominant term for small $s$ will be the one with $n=0$:
\eq
\f{-a}{\sinh 2a s}
\eqx
Taking the difference, as in (\ref{e.residues}), reproduces exactly the periodic answer (\ref{e.periodic}). Similarly, one can recover the same behaviour for $t=2\pi k +s$ for any integer $k$.

We see, therefore, that the periodicity in time is a simple consequence of the summation structure of the CFT expression~(\ref{e.isingwindings}).

\subsubsection{Gravity interpretation}
\label{s.interpretation}

Due to the clear matching of the expressions for individual Wilson lines with appropriate windings with the factors $a/\sinh(a(z+2\pi n))$, we can reinterpret the CFT formula (\ref{e.isingwindings}) as a prescription for allowed windings of holomorphic and antiholomorphic Wilson lines representing the CFT primary field in the bulk Chern-Simons formulation.

Thus going back to the Wilson line bulk interpretation, we obtain a very surprising behaviour. The windings of the holomorphic and antiholomorphic parts of the Wilson lines appear to be completely independent (as in Fig.~\ref{fig.mirror}b) up to a very loose ``topological'' coupling
\eq
n-n' = 0 \mod 2
\eqx
which only requires the same parity of the windings.

As one can view the Wilson line networks as a sort of first-quantized picture of matter fields, the independence of the holomorphic and antiholomorphic Wilson lines strongly suggests that the dual matter field is really a \emph{composite} of a field coupled only to $A$ and another field coupled only to $\Ab$. Those fields always couple in \emph{pairs} to the external sources (CFT operators inserted on the boundary) and to other pairs of fields corresponding to other CFT primary fields with a nonvanishing OPE coefficient.

As such, this may seem as a curiosity, a bizarre complication appearing in three dimensions. However, once we take into account that the two Chern-Simons theories \emph{only together} describe gravity, the separation of fields gets far reaching consequences. 

Indeed, the metric and also the black hole horizon is only a property of \emph{both} CS gauge fields $A$ and $\Ab$ considered jointly. A \emph{single} $SL(2,\RR)$ CS theory does not incorporate the causal structure of the bulk (recall section~\ref{s.3dcs}). The fields which seem to interact just with a single theory thus may effectively ignore key features of the black hole background --- leading to the re-emergence of the signal.
They thus evade the black hole information paradox.

As the metric and thus geometry is a derived and not fundamental concept in Chern-Simons gravity, there is \emph{a-priori} a possibility of matter fields interacting with the CS gauge fields in a manner which is not reducible to a conventional interaction with the metric.
To our surprise, the CFT holographic description seems to require exactly this to happen.

Yet, this violation is rather mild, as for small times one recovers conventional black-hole like behaviour such as exponential QNM damping as well as the plateau, which can be also understood as coming from a loop around the horizon.
It is only at the level of nontrivial winding that we see a modification of the conventional behaviour.

\subsubsection{Further comments}

We briefly address here two possible issues associated with the above hypothesis. Firstly, as we postulate that the composite matter fields interact e.g. only with the $A$ gauge field, one could think that this would be equivalent to the conventional interaction for a field with $h\neq 0$ and $\hb=0$. This is not the case, however, as $\hb=0$ is not a singlet representation of $SL(2, \RR)$. Indeed, for vanishing conformal weight, even though the prefactor in the Wilson line expression~(\ref{e.wilsongen}) goes away, the nontrivial Dirac delta remains.

Secondly, a similar periodicity appears for $T=0$. Its origin in that case is, however, quite different and does not involve the subtleties discussed here. Indeed let us calculate the Wilson line
\eq
\int \dl(x_f) W(z_f, x_f; z_i, x_i) \f{1}{x_i^{2h}} dx_i dx_f
\eqx
Since $\cor{T(z)}=-c/24$, the solution of (\ref{e.wb})-(\ref{e.wc}) is
\eq
x_T(z) = -2 \tan \f{z-z_f}{2}
\eqx
Plugging this in, specializing to $h=\hb=\f{1}{2}$ we obtain finally
\eq
\f{1}{2\sin \f{z_{fi}}{2}} \f{1}{2\sin \f{\zb_{fi}}{2}}
\eqx
Applying the residue formula yields
\eq
\label{e.zeroTR}
R_\eps(t) = \f{1}{\sin t}
\eqx
which is explicitly periodic, even though we did not include any winding.
In fact, as the bulk geometry is then $AdS_3$, there should not be any singularity\footnote{Although it would be interesting to see how this is realized in detail on the fully quantum level.} and therefore the winding modes should not be present at all, as well as the $\sg$-loop that we saw earlier. We note that (\ref{e.zeroTR}) does not have (as expected) any exponential damping whatsoever.

\red{
\subsubsection{Remarks on generality and the black hole information paradox}
\label{s.generality}

The evasion of the black hole information paradox which we observed above, came from the fact that at long timescales, the matter fields interact separately with the $A$ and $\Ab$ gauge fields, and hence do not feel the bulk causal structure.
Although the concrete calculation was performed for the Ising CFT, we believe that the overall conclusion should be of very general applicability in $AdS_3/CFT_2$. 

Indeed, the independent allowed windings of the holomorphic and antihomorphic Wilson lines do not seem to depend on any details specific to the Ising model CFT. The characteristics of the CFT operators only entered through the values of the conformal weights.
Hence we expect that the independent windings is a general property of the Chern-Simons matter dual to generic CFT operators.

Indeed, the recurrence which led to the independent windings came on the Ising CFT side from the following combination of torus 2-point blocks:
\eq
\twop{$\varepsilon$}{$\varepsilon$}{$\varepsilon$}{$1$}{\textbf{a}}\!\!\!\!(z) \cdot
\!\!\!\!\twop{$\varepsilon$}{$\varepsilon$}{$\varepsilon$}{$1$}{\textbf{a}}\!\!\!\!(\zb)
+
\twop{$\varepsilon$}{$\varepsilon$}{$1$}{$\varepsilon$}{\textbf{a}}\!\!\!\!(z) \cdot
\!\!\!\!\twop{$\varepsilon$}{$\varepsilon$}{$1$}{$\varepsilon$}{\textbf{a}}\!\!\!\!(\zb)
\eqx
Such a combination of torus conformal blocks is in fact universal and appears for arbitrary primary fields in any CFT:
\eq
\twop{$h$}{$h$}{$h$}{$0$}{\textbf{a}}\!\!\!\!(z) \cdot
\!\!\!\!\twop{$\hb$}{$\hb$}{$\hb$}{$0$}{\textbf{a}}\!\!\!\!(\zb)
+
\twop{$h$}{$h$}{$0$}{$h$}{\textbf{a}}\!\!\!\!(z) \cdot
\!\!\!\!\twop{$\hb$}{$\hb$}{$0$}{$\hb$}{\textbf{a}}\!\!\!\!(\zb)
\eqx
where we relabelled the fields by their conformal weights.
Each of these holomorphic and antiholomorphic blocks should have \emph{separate} simple periodic behaviour, up to a possible phase factor (c.f.~(\ref{e.blockperiodicity})).
On the dual Chern-Simons side, as periodicity is directly implemented through summation over windings, the \emph{independent} periodicity properties of the holomorphic and antiholomorphic conformal blocks similarly indicate that the bulk matter fields interact separately with the $A$ and $\Ab$ gauge fields, and hence do not feel the bulk causal structure, thus evading the black hole information paradox.

We note, however, that for generic conformal weights, the temporal evolution may be more involved as one would have branch cuts instead of poles which would significantly complicate obtaining the integrated retarded Green's function. We leave this problem for future investigations.

A direct consequence of the above reasoning is that the same mechanism should apply in the clearly semiclassical setting of $c\to \infty$. However, we would like to emphasize that in the high temperature limit in the Ising model CFT, our evaluation of the Chern-Simons Wilson lines was effectively semiclassical, as we substituted the thermal expectation values (\ref{e.Athermal}) for the Chern-Simons gauge fields which correspond to the classical BTZ black hole solution.

For generic CFTs, there are of course many more possible contributing two-point conformal blocks, in particular ones where the holomorphic and antiholomorphic blocks would be significantly different e.g.
\eq
\twop{$\varepsilon$}{$\varepsilon$}{$\varepsilon$}{$1$}{\textbf{a}}\!\!\!\!(z) \cdot
\!\!\!\!\twop{$\varepsilon$}{$\varepsilon$}{$1$}{$\varepsilon$}{\textbf{a}}\!\!\!\!(\zb)
\eqx
Such a contribution is forbidden in the Ising CFT, but could appear in the free fermion theory (without projecting out any states). There, it would correspond to $\eps$ splitting into $\psi$ and $\bar{\psi}$ and recombining again. We would therefore expect this and similar expressions in nondiagonal theories to contribute primarily to the plateau region (as in Fig.~\ref{fig.qnm}b). 
}

\subsection{The ``quantum'' burst}

Let us finally comment on the burst occurring at $t=\pi$ shown in Fig.~\ref{fig.tcsa}. One can check, that on the CFT side it comes completely from the part of the torus 2-point function involving the $\sg$ field i.e.
\eq
f^\sg_{\eps\sg\eps}(z=x-t) f^\sg_{\eps\sg\eps}(\zb=x+t)
\eqx
Therefore, on the bulk side, we expect it to come wholly from the $\sg$ loop around the singularity. Unfortunately, we are not able to verify this explicitly as for $t=\pi+s$ both arguments are small, so require nontrivial energy-momentum tensor contractions. Indeed, when computing the residue at $x=t$, we have
\eq
\label{e.burstres}
\res_{z=0} f^\sg_{\eps\sg\eps}(z) \cdot
f^\sg_{\eps\sg\eps}(2\pi+2s)
\eqx
But due to the periodicity
\eq
f^\sg_{\eps\sg\eps}(2\pi+z) = - f^\sg_{\eps\sg\eps}(z) \ ,
\eqx
the second factor in (\ref{e.burstres}) also has essentially a small argument.
The Ising CFT result in the high temperature limit is
\eq
R_\eps(\pi+s) \sim -\f{2\pi T\, e^{-\f{\pi^2}{2} T}}{1 + e^{-\f{\pi^2}{2} T}} \coth 2\pi T s
\eqx
It would be very interesting to reproduce this directly on the Chern-Simons side.
In any case, the specific 2-point torus block indicates that it is a quantum effect (a $\sg$ loop diagram around the horizon/singularity) so it is really a ``quantum burst''.

\section{Discussion}

In this paper we show that the main features of the temporal dynamics in the Ising model CFT at high temperatures can be understood in terms of a bulk gravitational Chern-Simons description with matter fields in a BTZ black hole background. We reproduce an exponential QNM-like damping, the magnitude of the subsequent plateau (``quantum halo'') -- which can be understood as coming from a loop diagram around the black hole singularity or horizon (if one would use the scalar field language as in~\cite{Kraus:2016nwo}).
We argue, however, that the scalar field formulation for the bulk matter dual to CFT primary fields is only approximate.

Indeed, a prominent feature of the CFT dynamics is a periodicity in time (with period $2\pi$, equal to the size of the spatial circle, for the concrete observable investigated in the present paper).
This behaviour has a very simple interpretation on the CFT side, due to the equal level spacing in Virasoro representations. It is very surprising, however,  from the dual gravitational black hole perspective, as it corresponds to the re-emergence of the whole intact signal, already long after it has been absorbed by the black hole through exponential QNM-like decay.

We show that the behaviour leading to the re-emergence of the signal has its origin in the way that the dual matter field couples to gravity in the Chern-Simons language. Instead of a single field coupling to both Chern-Simons gauge fields (as would be the case for a scalar field in a BTZ black hole background), the structure of the CFT formulas implies that one should rather consider a pair of fields, each coupled to a single gauge field\footnote{With just a loose constraint requiring same parity of the winding number of the two fields.}. Then, with this provision, we reproduce the recurrence/periodicity in the high temperature limit using the standard BTZ black hole background of the Chern-Simons gauge fields. 

Despite sounding like an obscure technical oddity, the above observed structure seems to have far-reaching physical implications on the possible nature of gravity.
Indeed, since each field of the pair couples just to a single Chern-Simons gauge field, it effectively ignores some features of the geometry as given by the metric, as the metric and consequently the horizon is necessarily a function of \emph{both} gauge fields (see e.g. discussion in section~\ref{s.3dcs}). This behaviour is in principle possible, as the metric is not a fundamental object in the Chern-Simons formulation but rather is a derived quantity. Consequently, the gravity+matter system effectively evades the black hole information paradox in this setting. Indeed, as shown in the present paper, one directly recovers the \emph{exact re-emergence} of the signal despite using the conventional BTZ black hole Chern-Simons solution. 

\red{We further gave arguments, that the above mechanism is a generic feature in $AdS_3/CFT_2$ contexts, realized through Wilson line matter including, but not restricted to semiclassical gravity.}

One may speculate, that a similar mechanism may also work in other formulations of gravity in higher dimensions, where the metric is not fundamental. Clearly this is a very interesting question for further investigation. 

One could have an issue whether one should really consider the pair of Chern-Simons theories as ``real gravity''. The point of view adopted in the present paper is to consider holography as a \emph{definition} of a particular gravity theory, and to try to extract its properties without any prior preconceptions.
Gravity is then identified with the sector of the bulk theory dual to the energy-momentum tensor of the boundary QFT.
Ultimately, the holographic definition provides for us examples of what gravity theories could in principle look like. Whether anything similar is realized in the real world is of course another issue.

As a further comment, the recurrence and thus evading the black hole information paradox in the present context is tied to the properties of specific matter fields and how they interact with gravity and not to some specific mechanism purely within gravity. In particular, in case we would have an ordinary scalar field, recurrence would be absent. The role of Chern-Simons gravity in evading the paradox is that since the metric is not fundamental, such nonstandard coupling is at all possible.
Surprisingly, the dual boundary CFT seems to require such coupling.

Note, however, that despite the differences, for small times the observed behaviour is just as one would expect from a conventional scalar field in a BTZ background --- with exponential damping and a ``quantum halo''. The nonstandard features, in the present context, make their impact only for long times.

This work leads to numerous questions for further study. In particular, it would be important to make the link from (quantum) $SL(2,\RR)$ Chern-Simons to the Bershadsky-Ooguri construction much more precise. Also, developing precise rules and techniques for the evaluation of Wilson line networks on the quantum level at finite temperature in order to have an exact matching with the known Ising CFT results would be very interesting. A spacetime $2^{nd}$-quantized formulation of the matter fields coupled to Chern-Simons would be very helpful for a deeper understanding of the observed phenomena and to investigate any possible alternative interpretations.

In the course of this work, we also observed quantum effects ``opening up'' a throat in place of the horizon in the quantum expectation value of the metric. The subsequent arguments, however, seem to point more to a rather ``metric-free'' picture. It would be interesting to investigate, nevertheless, to what extent the average geometry has any clear physical impact at the quantum level.

The present paper analyzed virtually the simplest possible CFT --- the Ising CFT. It would be interesting to investigate higher minimal models which have a much richer structure of primary fields\red{, as well as theories with non-diagonal modular invariants.}
Naturally, deviating from conformality, as well as understanding the impact of integrability in this case from the bulk perspective would be fascinating. 
In higher dimensionality, similar periodicity can also occur (e.g. for $\NN=4$ SYM on $S^3\times \RR$ at zero coupling), but then we lack even a framework of the corresponding gravitational picture. We hope to pursue some of these lines of investigation in the future.

\medskip

\noindent
{\bf Acknowledgements.} 
I would like to thank Zoltan Bajnok for initial collaboration on the TCSA. I benefitted from questions, answers and discussions with Zoltan Bajnok, David Horv{\'a}th, Veronica Hubeny, Shota Komatsu, Mukund Rangamani, Andrzej Rostworowski and Ashoke Sen.
An early part of this work was done during the KITP program {\it What is String Theory? Weaving Perspectives Together}, supported by grant NSF PHY-2309135 to the Kavli Institute for Theoretical Physics (KITP).
This work was supported in part by a Priority Research Area DigiWorld grant under the Strategic Program Excellence Initiative at the Jagiellonian University (Kraków, Poland).

\clearpage

\appendix

\section{The $T\Tb$ inequality}
\label{s.inequality}

Here we will provide some more general conditions when the inequality
\eq
\cor{T\Tb} \geq \cor{T}\cor{\Tb}
\eqx
discussed in section~\ref{s.fragility} holds, with equality occurring only for states of definite energy. 
Let us consider two cases: i) a density matrix diagonal in the basis of states with definite energy and zero momentum, 
and ii) a pure state with components in \emph{distinct} Virasoro representations
\eq
\ket{\Psi} = \sum c_i \ket{h_i, \hb_i=h_i}
\eqx
This assumption is necessary so that $\cor{\Psi|T(z)|\Psi}$ is independent of $z$ on the cylinder. On the other hand, $\ket{h_i, \hb_i=h_i}$ does not need to be a primary state.

In both cases, the inequality reduces to
\eq
\label{e.mathineq}
\sum_i p_i h_i^2 - \left( \sum_i p_i h_i \right)^2 \geq 0
\eqx
with $\sum_i p_i =1$.
Without loss of generality, we can assume all $h_i$ being distinct and $p_i>0$. Then with some effort, (\ref{e.mathineq}) follows by reinterpreting it as a statement of the positive definiteness of a quadratic form and following the proof of Sylvester's criterion with the caveat that the last determinant vanishes.

When $\hb_i \neq h_i$, the inequality does not hold in general (e.g. using $\ket{h,0}$ and $\ket{0,h}$), but we checked numerically that it holds for the thermal density matrix of the Ising CFT.

\section{Antiholomorphic Wilson line}
\label{s.antiholo}

As the $A$ and $\Ab$ gauge potentials for the BTZ black hole are not explicitly symmetric, here we comment why the outcome in both sectors has to be analogous.
For the $SL(2,\RR)$ generators appearing in $\Ab$, it is convenient to choose a different differential realization than (\ref{e.diffgen}), but with the same Casimir:
\eqn
\label{e.diffgenAb}
L^+ &=& -\xb^2 \partial_\xb -2 \hb \xb  \\
L^0 &=& -\xb\partial_\xb - \hb \\
L^- &=& -\partial_\xb
\eqnx
With this choice, the $-\Ab$ given by (\ref{e.Ab}) and $A$ given by~(\ref{e.A}) are essentially identical as differential operators in the $x/\xb$ variables. Hence, the computation of the Wilson line $e^{-\int\Ab}$ matrix element will be identical with~(\ref{e.wilsongen}) in the main text. 

It remains to discuss the external primary states. For that we have to compare the bulk $SL(2,\RR)$ algebra with the CFT properties in both the holomorphic and antiholomorphic sectors. Deriving the transformation laws of $T(z)$ and $\Tb(\zb)$ from bulk gauge transformations 
\eq
\dl A = d\lm + [A, \lm] \qq \dl \Ab = d\lmb + [\Ab, \lmb]
\eqx
with $\lm = \lm_a(z) L^a$ one gets
\eqn
\dl T\!\! &=&\!\! \lm_+(z) \partial T(z) + 2 \partial\lm_+(z) T(z) -\f{c}{12} \partial^3 \lm_+(z) \nonumber\\
\dl \Tb\!\! &=&\!\!  -\lmb_-(\zb) \partialb \Tb(\zb) - 2 \partialb\lmb_-(\zb) \Tb(\zb) +\f{c}{12} \partialb^3 \lmb_-(\zb) \nonumber
\eqnx
Since $-\lmb_-(\zb)$ plays the same role as $\lm_+(z)$, this indicates that one should perform the following $SL(2,\RR)$ automorphism on the anitholomorphic generators
\eq
L^\pm \to  - L^\mp \qq L^0 \to -L^0
\eqx
Therefore, the condition for the primary state will take exactly the same form as in the holomorphic sector leading to
\eq
\ket{\hb} = \f{1}{\xb^{2\hb}}
\eqx
Hence the antiholomorphic computation will be an exact parallel of the holomorphic one.

\section{Connected TT correlator formulas}
\label{s.tt}

The connected correlation function on the torus $\cor{T(z) T(z')}_T - \cor{T(z) }_T \cor{T(z')}_T$ has been computed in~\cite{DiFrancesco:1987ez}. It is expressed as
\eq
\sum_{\nu=2}^4 \f{Z_\nu}{Z_{ising}} \biggl[ \cor{T(z) T(z')}_\nu - \cor{T(z) }_\nu \cor{T(z')}_\nu \biggr]
\eqx
where
\eq
\f{Z_\nu}{Z_{ising}} = \f{\th_\nu(0)}{\th_2(0)+\th_3(0)+\th_4(0)}
\eqx
The connected correlator in each sector $\nu$ is expressed\footnote{The factor $\f{1}{2}$ and the argument $z/2$ in this formula comes from using Mathematica conventions and $2\pi$ spatial periodicity of the torus.} in terms of 
\eq
\PP_\nu(z) = \f{1}{2} \f{\th_1'(0)}{\th_\nu(0)} \f{\th_\nu(\f{z}{2})}{\th_1(\f{z}{2})}
\eqx
as
\eq
\cor{T(z) T(z')}_\nu^{conn} = -\f{1}{4}\left( \PP_\nu'(z)^2 - \PP_\nu''(z) \PP_\nu(z) \right)
\eqx

\section{Ising CFT 2-point a-cycle torus blocks}
\label{s.torusblocks}

The 2-point torus blocks for the Ising CFT are given by the following formulas\footnote{Here we correct for a mislabelling in eq. (12.94) in ref.~\cite{DiFrancesco:1997nk}, and adjust the formulas to work with Mathematica conventions and spatial periodicity $2\pi$.}:
\eqn
\twop{$\varepsilon$}{$\varepsilon$}{$\varepsilon$}{$1$}{}\!\!\!\!\!\! &=&\!\!
\f{1}{4\sqrt{\eta(\tau)}}\, \f{\th'_1(0)}{\th_1(\f{z}{2})} 
\left[ \f{\th_3(\f{z}{2})}{\sqrt{\th_3(0)}} + \f{\th_4(\f{z}{2})}{\sqrt{\th_4(0)}}\right] \nonumber\\
\twop{$\varepsilon$}{$\varepsilon$}{$1$}{$\varepsilon$}{}\!\!\!\!\!\! &=&\!\!
\f{1}{4\sqrt{\eta(\tau)}}\, \f{\th'_1(0)}{\th_1(\f{z}{2})} 
\left[ \f{\th_3(\f{z}{2})}{\sqrt{\th_3(0)}} - \f{\th_4(\f{z}{2})}{\sqrt{\th_4(0)}}\right] \nonumber\\
\twop{$\varepsilon$}{$\varepsilon$}{$\sg$}{$\sg$}{}\!\!\!\!\!\! &=&\!\!
\f{1}{2\sqrt{2\eta(\tau)}}\, \f{\th'_1(0)}{\th_1(\f{z}{2})} 
 \f{\th_2(\f{z}{2})}{\sqrt{\th_2(0)}} 
\eqnx
where $0$ and $z$ are the insertion points of the $\eps$ operators.
In these formulas, the loop represents the \textbf{b}-cycle of the torus, and in the $z \to 0$ limit, these formulas become proportional to the characters of the representation marked on the lower arc of the loop.

In order to pass to \textbf{a}-cycle blocks, we use the modular transformation matrix:
\eqn
\twop{$\varepsilon$}{$\varepsilon$}{$\varepsilon$}{$1$}{\textbf{a}}\!\!\!\!\!\! &=& 
\f{1}{2}\!\!\! \twop{$\varepsilon$}{$\varepsilon$}{$\varepsilon$}{$1$}{}
+
\f{1}{2}\!\!\! \twop{$\varepsilon$}{$\varepsilon$}{$1$}{$\varepsilon$}{}
+
\f{\sqrt{2}}{2}\!\! \twop{$\varepsilon$}{$\varepsilon$}{$\sg$}{$\sg$}{}
\nonumber\\
\twop{$\varepsilon$}{$\varepsilon$}{$1$}{$\varepsilon$}{\textbf{a}}\!\!\!\!\!\! &=& 
\f{1}{2}\!\!\! \twop{$\varepsilon$}{$\varepsilon$}{$\varepsilon$}{$1$}{}
+
\f{1}{2}\!\!\! \twop{$\varepsilon$}{$\varepsilon$}{$1$}{$\varepsilon$}{}
-
\f{\sqrt{2}}{2}\!\! \twop{$\varepsilon$}{$\varepsilon$}{$\sg$}{$\sg$}{}
\nonumber\\
\twop{$\varepsilon$}{$\varepsilon$}{$\sg$}{$\sg$}{\textbf{a}}\!\!\!\!\!\! &=& 
\f{\sqrt{2}}{2}\!\! \twop{$\varepsilon$}{$\varepsilon$}{$\varepsilon$}{$1$}{}
-
\f{\sqrt{2}}{2}\!\! \twop{$\varepsilon$}{$\varepsilon$}{$1$}{$\varepsilon$}{}
\eqnx\\
which leads to the final results:
\eqn
\twop{$\varepsilon$}{$\varepsilon$}{$\varepsilon$}{$1$}{\textbf{a}}\!\!\!\!\!\! &=&\!\!
\f{1}{4\sqrt{\eta(\tau)}}\, \f{\th'_1(0)}{\th_1(\f{z}{2})} 
\left[ \f{\th_3(\f{z}{2})}{\sqrt{\th_3(0)}} + \f{\th_2(\f{z}{2})}{\sqrt{\th_2(0)}}\right] \nonumber\\
\twop{$\varepsilon$}{$\varepsilon$}{$1$}{$\varepsilon$}{\textbf{a}}\!\!\!\!\!\! &=&\!\!
\f{1}{4\sqrt{\eta(\tau)}}\, \f{\th'_1(0)}{\th_1(\f{z}{2})} 
\left[ \f{\th_3(\f{z}{2})}{\sqrt{\th_3(0)}} - \f{\th_2(\f{z}{2})}{\sqrt{\th_2(0)}}\right] \nonumber\\
\twop{$\varepsilon$}{$\varepsilon$}{$\sg$}{$\sg$}{\textbf{a}}\!\!\!\!\!\! &=&\!\!
\f{1}{2\sqrt{2\eta(\tau)}}\, \f{\th'_1(0)}{\th_1(\f{z}{2})} 
 \f{\th_4(\f{z}{2})}{\sqrt{\th_4(0)}} 
\eqnx
with $\th_2$ and $\th_4$ interchanged.
The constant plateau value arises from the high temperature asymptotics of $\th_4$.

\vspace{10cm}

\clearpage

\bibliography{references}

\end{document}